\documentclass[twocolumn, prx, superscriptaddress,notitlepage]{revtex4-1}
\usepackage{graphicx}
\usepackage{dcolumn}
\usepackage{bm}
\usepackage[most]{tcolorbox}
\usepackage{multirow}
\usepackage{gensymb}
\usepackage{CJK}

\usepackage[colorlinks, linkcolor=blue,anchorcolor=blue,citecolor=blue,urlcolor=blue]{hyperref}

\usepackage{color,soul}
\begin{document}
\begin{CJK*}{UTF8}{}
\title{Observation of high-order Mollow triplet by quantum mode control with concatenated continuous driving}

\author{Guoqing Wang \CJKfamily{gbsn}(王国庆)}
\affiliation{
   Research Laboratory of Electronics and Department of Nuclear Science and Engineering, Massachusetts Institute of Technology, Cambridge, MA 02139, USA}
\author{Yi-Xiang Liu \CJKfamily{gbsn}(刘仪襄)}
\affiliation{
   Research Laboratory of Electronics and Department of Nuclear Science and Engineering, Massachusetts Institute of Technology, Cambridge, MA 02139, USA}
\author{Paola Cappellaro}\email[]{pcappell@mit.edu}
\affiliation{
   Research Laboratory of Electronics and Department of Nuclear Science and Engineering, Massachusetts Institute of Technology, Cambridge, MA 02139, USA}
\affiliation{Department of Physics, Massachusetts Institute of Technology, Cambridge, MA 02139, USA}

\begin{abstract}
The Mollow triplet is a fundamental signature of quantum optics, and has been observed in numerous quantum systems. Although it arises in the ``strong driving''  regime of the quantized field, where the atoms undergo  coherent oscillations, it can be typically analyzed within the rotating wave approximation.  
Here we report the first observation of high-order effects in the Mollow triplet structure due to strong driving. 
In  experiments, we  explore the regime beyond the rotating wave approximation using concatenated continuous driving that has less stringent requirements on the driving field power.
We are then able to reveal additional transition frequencies, shifts in energy levels, and corrections to the transition amplitudes. In particular, we find that these amplitudes are more sensitive to high-order effects than the frequency shifts, and that they still require an accurate determination in order to achieve high-fidelity quantum control.  The experimental results are validated by Floquet theory, which enables the precise numerical simulation of the evolution and further provides an analytical form for an effective Hamiltonian that approximately predicts the spin dynamics beyond the rotating wave approximation.
\end{abstract}

\maketitle
\end{CJK*}	

\section{Introduction}

The Mollow triplet  was originally observed in coupled atom-laser systems, where the laser field creates a series of equidistant energy level manifolds and induces the coupling inside each manifold \cite{mollowPowerSpectrumLight1969}.  The coherent atomic oscillations modulate the laser driving field and spontaneous emission acts  as a detection tool for the triplet energy levels~\cite{Cohen-Tannoudji1996}. 
The Mollow triplet structure has been observed in atomic beams \cite{schramaIntensityCorrelationsComponents1992}, ions \cite{stalgiesSpectrumSingleatomResonance1996}, single molecules \cite{wriggeEfficientCouplingPhotons2008}, quantum dots \cite{nickvamivakasSpinresolvedQuantumdotResonance2009,flaggResonantlyDrivenCoherent2009,peirisTwocolorPhotonCorrelations2015,lagoudakisObservationMollowTriplets2017}, superconducting qubits \cite{baurMeasurementAutlerTownesMollow2009}, and cold atoms \cite{ortiz-gutierrezMollowTripletCold2019}. Its potential applications, such as heralded single-photon sources \cite{ulhaqCascadedSinglephotonEmission2012}, quantum sensing \cite{joasQuantumSensingWeak2017,starkNarrowbandwidthSensingHighfrequency2017,honigl-decrinisTwoLevelSystemQuantum2020}, and spin noise characterization \cite{poshakinskiySpinNoiseElectron2020}, make it a versatile tool in physics.  

Even an electromagnetic field in the microwave range, such as in magnetic resonance experiments, can produce a similar level structure. Typical Larmor precession under a single (classical) microwave driving (Rabi oscillation) is not able to reveal the Mollow triplet since the observation operator $\langle\sigma_z\rangle$ commmutes with the static splitting $\frac{\omega_0}{2}\sigma_z$. By applying multiple driving fields to orient the effective static splitting Hamiltonian along a transverse direction, it becomes possible to observe the Mollow triplet by Rabi oscillation, as observed in 
nitrogen-vacancy (NV) centers in diamond \cite{rohrSynchronizingDynamicsSingle2014,teissierHybridContinuousDynamical2017,pigeauObservationPhononicMollow2015}. 

In  previous experiments \cite{schramaIntensityCorrelationsComponents1992,stalgiesSpectrumSingleatomResonance1996,wriggeEfficientCouplingPhotons2008,nickvamivakasSpinresolvedQuantumdotResonance2009,flaggResonantlyDrivenCoherent2009,peirisTwocolorPhotonCorrelations2015,lagoudakisObservationMollowTriplets2017,baurMeasurementAutlerTownesMollow2009,ortiz-gutierrezMollowTripletCold2019,rohrSynchronizingDynamicsSingle2014,teissierHybridContinuousDynamical2017,pigeauObservationPhononicMollow2015} that showed the Mollow triplet, the driving strength was smaller than the static splitting, allowing one to conveniently solve the dynamics using the rotating wave approximation (RWA). Previous theoretical studies beyond the RWA have predicted frequency shifts and imbalanced sidebands \cite{LuBlochSiegertPhysRevA.86.023831,YanBlochMollowPhysRevA.88.053821}, yet experimental verification is still lacking due to the limited driving strength. The aim of this paper is to experimentally explore the Mollow triplet structure in the regime beyond the RWA. 

To overcome the usual constraints on the driving power needed to achieve the strong driving regime beyond the RWA, we use concatenated continuous driving (CCD). CCD is a continuous dynamical decoupling technique with multiple resonant modulated fields, and has been studied before in the context of qubit coherence protection  \cite{caiRobustDynamicalDecoupling2012,khanejaUltraBroadbandNMR2016,saikoSuppressionElectronSpin2018,cohenContinuousDynamicalDecoupling2017,farfurnikExperimentalRealizationTimedependent2017,rohrSynchronizingDynamicsSingle2014,laytonRabiResonanceSpin2014,saikoMultiphotonTransitionsRabi2015,teissierHybridContinuousDynamical2017,bertainaExperimentalProtectionQubit2020,caoProtectingQuantumSpin2020,wangCoherenceProtectionDecay2020}. As we show below, in the first interaction picture, the CCD driving term generates a Hamiltonian describing the typical AC driving of a two-level system, although with the static field along the x direction 
\cite{rohrSynchronizingDynamicsSingle2014,teissierHybridContinuousDynamical2017,pigeauObservationPhononicMollow2015,wangCoherenceProtectionDecay2020}. Then, since the static energy in the first interaction picture is set by the Rabi frequency, the ratio of the driving to the static splitting can be made much larger than in typical Rabi experiments \cite{cohenContinuousDynamicalDecoupling2017,farfurnikExperimentalRealizationTimedependent2017,Laucht16,wangCoherenceProtectionDecay2020}. In addition, the initial state does not commute with the interaction picture transformation, which makes it possible to observe the Mollow triplet by Rabi oscillation. Thus, CCD is a good experimental tool to study novel physics beyond the RWA, and in particular, the Mollow triplet. 

To analyze our experimental results, we use Floquet theory as a precise numerical tool to solve the evolution as a summation of a series of modes. To gain further insight, we also use Floquet theory as an analytical way to evaluate the effective Hamiltonian and calculate the corrections to frequency values and transition amplitudes resulting from the strong driving. We can thus provide a complete picture of the Mollow triplet structure beyond the rotating  wave approximation.

\section{Mollow triplet with CCD}
\label{sec:CCD}
\begin{figure}[h]
\includegraphics[width=86mm]{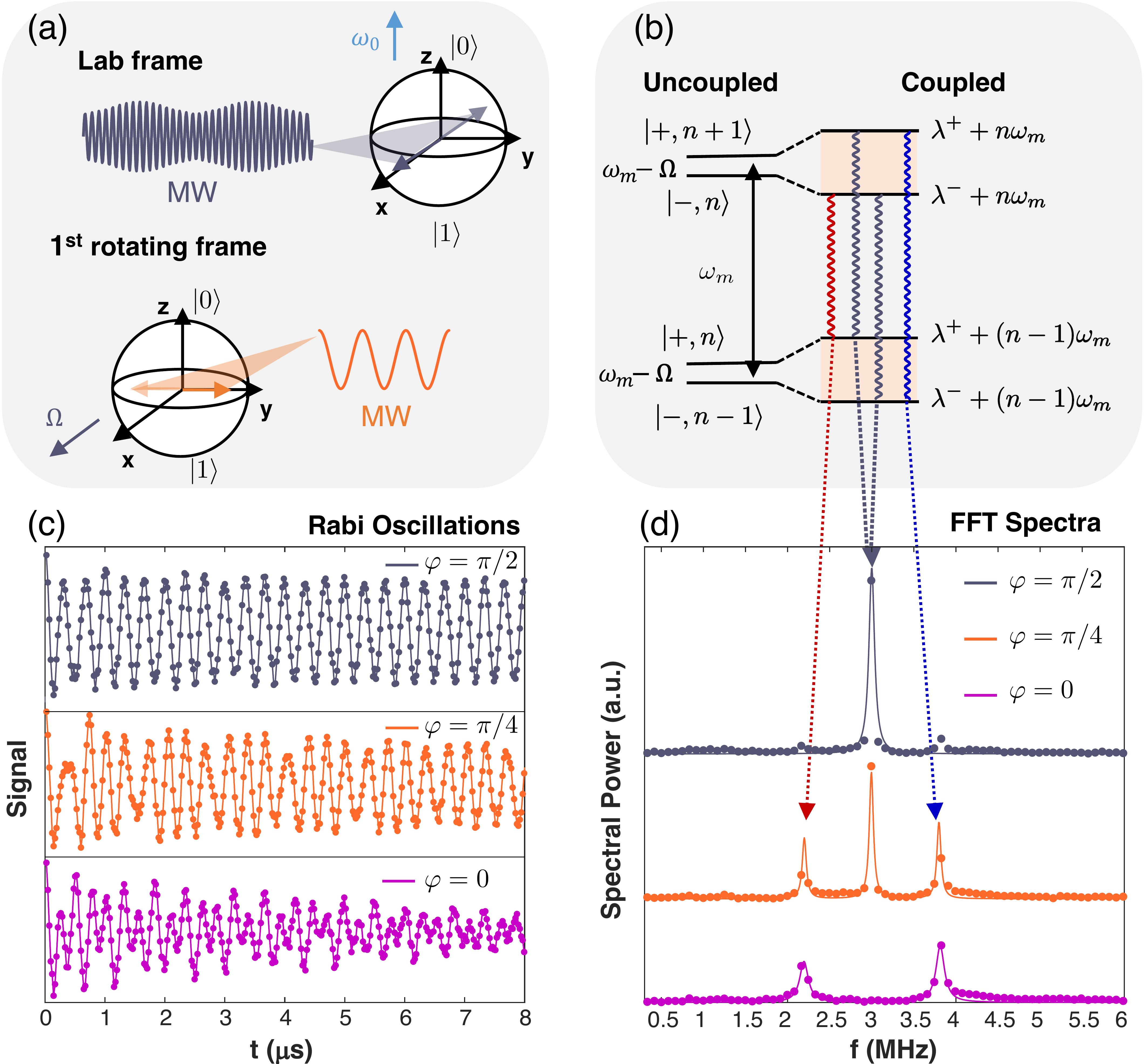}
\caption{\label{MollowPlot} \textbf{Observation of Mollow triplet with the CCD scheme.} \textbf{(a)} Principle of  amplitude-modulated CCD method. \textbf{(b)} Energy structure in the dressed atom picture explaining the origin  of the Mollow triplet structure. \textbf{(c)} Measured Rabi oscillations under three modulation phases $\phi=\pi/2$ (top), $\phi=\pi/4$ (middle) and $\phi=0$ (bottom). \textbf{(d)} Mollow triplet. Fourier spectra of the Rabi oscillations shown in (c). When $\phi=\pi/2$, only the center band exists, which corresponds to spin-locking. When $\phi=0$, only the sidebands exist. For other angles,  we can observe the three frequency components comprising the triplet structure.}
\end{figure}

Here we introduce the CCD scheme and show how it can be used to reveal the Mollow triplet. We study a two-level system with a static splitting $\omega_0$ along z, coupled to an amplitude-modulated microwave applied along the x axis given by $\Omega\cos(\omega t)-2\epsilon_m\sin(\omega t)\cos(\omega_m t+\phi)$. 
When the RWA condition $\Omega,\epsilon_m\ll\omega_0$ is satisfied, by going into the first rotating frame defined by $H_0=\frac{\omega}{2}\sigma_z$ and neglecting the counter-rotating term, the Hamiltonian becomes
\begin{equation}
    H_I=-\frac{\delta}{2}\sigma_z+\frac{\Omega}{2}\sigma_x+\epsilon_m\cos(\omega_mt+\phi)\sigma_y.
    \label{HI_AmpMod}
\end{equation} 
where $\delta=\omega-\omega_0$ is the frequency detuning.
Figure~\ref{MollowPlot}(a) shows the static and oscillating fields in the lab  and  first rotating frames. 
A similar Hamiltonian can also be engineered by phase modulation, where the phase of the microwave is modulated as $\Omega\cos\left(\omega t+\frac{2\epsilon_m}{\Omega}\cos(\omega_m t+\phi)\right)$ and the Hamiltonian in the interaction picture becomes 
\begin{equation}
H_I=-\frac{\delta}{2}\sigma_z+\frac{\Omega}{2}\sigma_x+\epsilon_m\frac{\omega_m}{\Omega}\sin(\omega_mt+\phi)\sigma_z.
    \label{HI_FreqMod}
\end{equation}

Evolution under the CCD Hamiltonian in Eq.~\eqref{HI_AmpMod} can be solved with a second interaction picture defined by $\frac{\omega_m}{2}\sigma_{x^{'}}$  with $\sigma_{x^{'}}=\frac{\Omega}{\Omega_R}\sigma_x-\frac{\delta}{\Omega_R}\sigma_z$ and $\Omega_R=\sqrt{\Omega^2+\delta^2}$. When $\delta=0$ and  $\epsilon_m\ll\Omega$, by dropping the counter-rotating terms, the Hamiltonian in the second interaction picture becomes 
\begin{equation}
H_I^{(2)}=\frac{\Omega-\omega_m}{2}\sigma_x+\frac{\epsilon_m}{2}(\cos\phi\, \sigma_y +\sin \phi\,\sigma_z).
\end{equation}
 The spin evolution in the first rotating frame is simply 
$|\psi(t)\rangle_I^{(1)}=e^{-i \frac{\omega_m}{2} \sigma_x t}e^{-i H_I^{(2)}t} |\psi(0)\rangle$, where $\omega_m$ modulates the dynamics happening in the second rotating frame. Then, we expect that the population in $|0\rangle$ to be generally a sum over the three frequencies $\omega_m$, $\omega_m\pm\Delta\lambda$, where $\Delta\lambda=\sqrt{\epsilon_m^2+(\omega_m-\Omega)^2}$ in the RWA (see Supplemental Material). Since the $\langle\sigma_z\rangle$ population measurement is along a direction that does not commute with the second interaction picture transformation along $\sigma_x$, the Mollow triplet structure can be revealed by the CCD scheme \cite{rohrSynchronizingDynamicsSingle2014,teissierHybridContinuousDynamical2017,pigeauObservationPhononicMollow2015}.

While this standard RWA description provides a simple analytical solution to the dynamics, using Floquet theory (see Appendix \ref{Derivation: Floquet}) provides further insight about the energy level structure, in analogy to the quantized field picture typically used to analyze the Mollow triplet in atomic physics.
Similar to the dressed atom approach \cite{Cohen-Tannoudji1996} where a series of equidistant level manifolds form the energy structure (see Fig.~\ref{MollowPlot}(b)), Floquet theory also predicts a series of energy manifolds~\cite{chuFloquetTheoremGeneralized2004}, $\cdots,\lambda^\pm-\omega_m,\lambda^\pm,\lambda^\pm+\omega_m,\cdots$, separated by integer-spacings of $\omega_m$, where $\lambda^\pm$ are the eigenenergies obtained by solving the Floquet Hamiltonian $H_F$. 
Within the RWA, the Floquet Hamiltonian is block-diagonal \cite{shirleySolutionSchrodingerEquation1965}. 
The off-diagonal terms in each block induce transitions between two adjacent level manifolds separated by $\omega_m$.  The system-field coupling  modifies the splitting by  $\Delta\lambda=\lambda^+-\lambda^-$, which creates the Mollow triplet structure with a center band at $\omega_m$ and two sidebands at $\omega_m\pm (\lambda^+-\lambda^-)$. 
The dressed state energy difference within the RWA is  $\lambda^+-\lambda^-=\sqrt{\epsilon_m^2+(\omega_m-\Omega_R)^2}$, where $\Omega_R=\sqrt{\Omega^2+\delta^2}$ is the effective Rabi frequency. 
Going beyond the RWA allows one to observe a richer framework of dynamics that we explore in the following. 

Our device is based on NV ensembles used in Ref.~\cite{jaskulaPhysRevApplied.11.054010} with $\sim10^{10}$ spins being addressed simultaneously. Two NV electronic spin states $|m_s=0\rangle$ and $|m_s=-1\rangle$ with a $2.207\text{GHz}$ splitting are used as the logical $|0\rangle$ and $|1\rangle$. An arbitrary waveform generator creates the modulated waveform to engineer the CCD Hamiltonian. By driving the NV centers  with the amplitude-modulated CCD scheme, we observe the Mollow triplet  in the Fourier spectra of the NV Rabi oscillations (see Figs.~\ref{MollowPlot}(c) and \ref{MollowPlot}(d)). In particular, we have phase control over the waveform, can separately observe the center band (spin-locking condition) as well as the sidebands, and carefully investigate their behavior.

\section{High-order Mollow triplet}
\label{sec:HigherOrder}
Previous work has explored  novel phenomena caused by strong driving, such as the Bloch-Siegert shift~\cite{BlochSiegertOldpaperPhysRev.57.522,ahmadTheoryBlochSiegertShift1974}, Landau-Zener-St\"{u}ckelberg interference \cite{shevchenkoLandauZenerStuckelberg2010,huangLandauZenerStuckelbergInterferometrySingle2011}, coherent destruction of tunneling (CDT) \cite{grossmannCoherentDestructionTunneling1991,grossmannCoherentTransportPeriodically1993}, novel behaviors in the quantum Zeno and anti-Zeno effects \cite{zhengQuantumZenoAntiZeno2008,aiQuantumAntiZenoEffect2010}, and asymmetric spectral features of the Mollow triplet \cite{YanBlochMollowPhysRevA.88.053821}. 
Here we theoretically and experimentally study the high-order Mollow triplet structure induced by strong driving, including corrections to energy levels and transition amplitudes.

To analyze our experimental results, we use Floquet theory which is known to accurately predict the system dynamics, including high-order phenomena, due to the counter-rotating terms beyond the RWA.
%
Floquet theory describes the periodic Hamiltonian as a time-independent operator $H_F$ in Fourier space. One can then numerically calculate the  evolution  by truncating  the (infinite) Floquet matrix \cite{leskesFloquetTheorySolidstate2010,shirleySolutionSchrodingerEquation1965}. Due to the simplicity of the computation, we perform a truncation to the Floquet matrix to high order ($400\times 400$ blocks) to accurately calculate the evolution. Analytical approximations in various parameter ranges are also possible, either by  moving to a suitable interaction picture where on-resonance terms highlight the corrections to the Hamiltonian due to the counter-rotating terms  \cite{ashhabTwolevelSystemsDriven2007,YanBlochSiegertPhysRevA.91.053834,YanBlochSiegertDisspationPhysRevA.90.053850,YanBlochMollowPhysRevA.88.053821,LuBlochSiegertPhysRevA.86.023831,zhouObservationTimeDomainRabi2014},
or by iteratively block-diagonalizing the time-independent Floquet Hamiltonian with a unitary transformation \cite{leskesFloquetTheorySolidstate2010}. 
In addition to providing numerical and approximate solution model, the Floquet picture, with its ladder-like energy structure, also provides an insightful understanding of the dynamics, in analogy to transitions and hoppings between energy levels in the dressed atom approach \cite{Cohen-Tannoudji1996,chuFloquetTheoremGeneralized2004}. Floquet theory can then  be  used as a tool to engineer desired Hamiltonian interactions, with applications relevant for areas from many-body systems to quantum control \cite{eckardtColloquiumAtomicQuantum2017,Childs9456}.

When applied to a two-level system, Floquet theory predicts a series of energy level manifolds yielding three types of transition frequencies (\textit{modes}),  $n\omega_m$, $n\omega_m\pm(\lambda^+-\lambda^-)$ (see Appendix.~\ref{Derivation: Floquet}). The exact evolution is a superposition of two Floquet eigenstates $c^+\Psi^+(t)+c^-\Psi^-(t)$ and the ensuing Rabi oscillation is a sum over these components, $P_{|0\rangle}(t) = \sum_{i,n} |a_{i,n}|\cos(\omega_{i,n} t+\phi_{i,n})$ where ${i,n}$ denotes the $n$th triplet with $i=-1,0,1$, and the transition amplitudes $a_{i,n} = |a_{i,n}|\exp(i\phi_{i,n})$ are obtained from Floquet theory. 
$|a_{i,n}|$ can be tuned by varying the initial conditions and the driving phase. This allows for control over the frequency modes involved in the evolution (\textit{mode control}).

A single mode evolution with only frequency components $n\omega_m$ can be achieved when the initial state is one of the two eigenstates $\Psi^\pm(0)$ and $c^+c^-=0$, which is equivalent to a spin-locking condition. In the RWA case, only components with $n=1$ can be observed in the Mollow triplet. However, when the counter-rotating terms in the Floquet matrix are not negligible, frequency components with $n>1$ emerge  (see Appendix \ref{Derivation: Floquet}). At the same time, values of the energy level $\lambda^\pm$ and transition amplitudes $|a_{i,n}|$ will both deviate from the RWA predictions.

In the following, we analyze the corrections to the energy levels and transition amplitudes. The Floquet approach can be used to analytically derive an effective, approximate Hamiltonian by systematically including higher order corrections beyond the RWA. By applying the van Vleck transformation \cite{leskesFloquetTheorySolidstate2010}, a unitary transformation to block-diagonalize the Floquet Hamiltonian, we obtain the effective Hamiltonian up to first order correction in the second rotating frame \begin{equation}
    \label{HIeff}
    H_{I,\text{eff}}^{(2)}=\frac{\epsilon_m}{2}\sigma_y+\frac{\epsilon_m^2}{8\Omega}\sigma_x+\frac{\Omega-\omega_m}{2}\sigma_x
\end{equation} 
where $\frac{\epsilon_m^2}{8\Omega}\sigma_x$ is the correction term and we assumed $\delta=0$ so that $\Omega_R=\Omega$ and the correction term is along x. 
Under the resonance condition $\omega_m=\Omega$, the shift of the level splitting inside each energy manifold is $\delta\epsilon_m=2\sqrt{(\frac{\epsilon_m}{2})^2+(\frac{\epsilon_m^2}{8\Omega})^2}-\epsilon_m=\frac{\epsilon_m^3}{32\Omega^2}$ with $\delta\epsilon_m/\epsilon_m=\mathcal{O}((\frac{\epsilon_m}{\Omega})^2)$ which quadratically depends on the ratio $\epsilon_m/\Omega$. 
The direction of the effective microwave field deviates from the y axis by a small angle $\delta\varphi\approx\frac{\epsilon_m^2}{8\Omega}/\frac{\epsilon_m}{2}=\mathcal{O}(\frac{\epsilon_m}{\Omega})$ within the XY plane. 
Such a linear dependence on  $\frac{\epsilon_m}{\Omega}$ induces drastic changes of the transition amplitudes $|a_{i,n}|$. 
The amplitude of each frequency component can be found analytically using the Hamiltonian in Eq.~\eqref{HIeff}, yielding a shift with respect to the RWA results. 
The amplitude shift of the $i$th frequency component, $\delta|a_{i,n}|$, depends on the initial state $|\Psi(0)\rangle$:  
for the central band, $\delta|a_{0,1}|=0$ when $|\Psi(0)\rangle$ is perpendicular to both the driving direction y and the Hamiltonian correction direction x; 
$\delta|a_{0,1}|\approx\frac{\sqrt{\pi}}{32\sqrt{2}}(\frac{\epsilon_m}{\Omega})^2=\mathcal{O}((\frac{\epsilon_m}{\Omega})^2)$ when $|\Psi(0)\rangle$ is along the driving direction y; 
and $\delta|a_{0,1}|\approx\frac{\sqrt{\pi}}{8\sqrt{2}}\frac{\epsilon_m}{\Omega}=\mathcal{O}(\frac{\epsilon_m}{\Omega})$ when $|\Psi(0)\rangle$ is along x. 
For the sidebands, in all three scenarios, the shifts are linear in $\frac{\epsilon_m}{\Omega}$,  $\delta|a_{\pm 1,1}|\approx \pm\frac{\sqrt{\pi}}{16\sqrt{2}}\frac{\epsilon_m}{\Omega}$ with $\pm$ corresponding to the two sidebands. 
Ultimately, the existence of the correction term breaks the symmetry of the two sidebands under the resonant condition $\omega_m=\Omega$ and results in large changes of the transition amplitudes especially for the sidebands. In comparison, the energy level correction is less sensitive to the counter-rotating effects.

\subsection{Sideband asymmetry}
\begin{figure}[b]
\includegraphics[width=86mm]{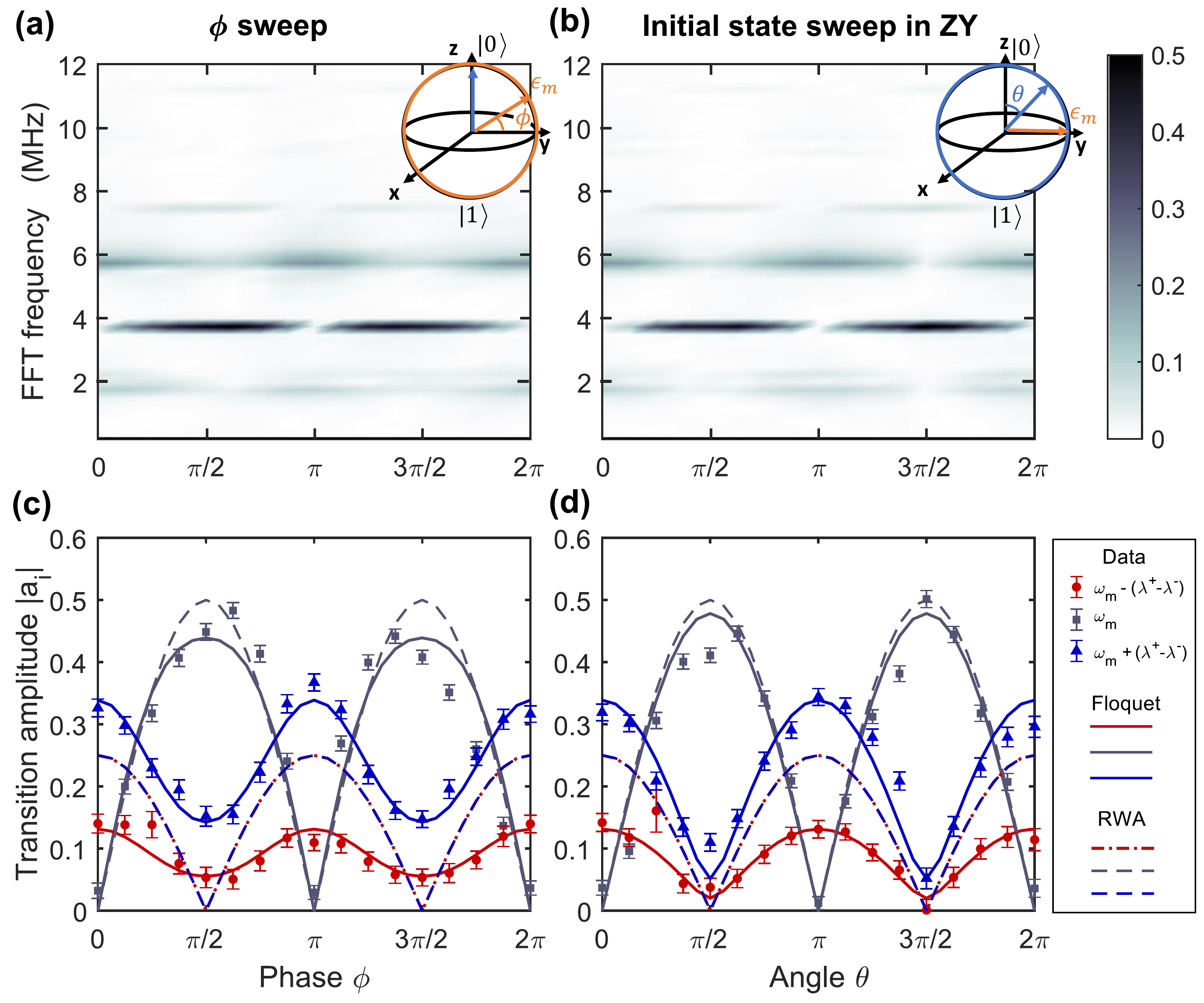}
\caption{
\label{phasestatesweep}  
\textbf{Sideband asymmetry of the Mollow triplet revealed by  mode evolution control.} 
Top panels (a-b): Fourier spectra of the Rabi oscillations. Bottom panels (c-d): transition amplitudes $|a_{i,1}|$. 
Parameters are $\Omega=\omega_m=(2\pi)3.75\text{MHz},\epsilon_m=(2\pi)2.08\text{MHz}$. In (a)(c) we sweep the phase of the drive, while in (b-d) the phase of the initial state.  
(a)(c) The initial state is  $|0\rangle$ and then the amplitude-modulated CCD waveform is applied. Rabi oscillations are measured from 0 to $4\mu s$ under different phases $\phi$ of the modulation term $\epsilon_m\cos(\omega_m t+\phi)$. The transition amplitudes $|a_{i,1}|$ are extracted by fitting the Rabi oscillation to $a_0+\sum_{i=1}^{3}a_i e^{-t^2/\tau_i^2}\cos(\omega_i t+\phi_i)$. (b)(d) The initial state is $\cos(\frac{\theta}{2})|0\rangle+i\sin(\frac{\theta}{2})|1\rangle$ and the CCD drive is applied with fixed phase, $\phi=0$. We plot the Rabi Fourier spectra and transition amplitudes following the same procedure as for (a)(c).
}
\end{figure}
We implement amplitude-modulated CCD experiments to systematically study the evolution mode control under the resonance condition $\delta=0$, with $\Omega=\omega_m=(2\pi)3.75\text{MHz}$. 
By applying a strong oscillating field, $\epsilon_m=(2\pi)2.08\text{MHz}$,  
we are able to observe the sideband asymmetry and the eigenstate shifts predicted by the theoretical analysis above. 
We implement mode control by sweeping the phase $\phi$ of the driving, or the angle $\theta$ of the initial states in the ZY plane (see also Appendix Fig.~\ref{phasestatesweep_Appendix} for results obtained when sweeping the initial state direction in the XY and XZ planes). 
The corresponding Rabi oscillations are measured by projecting the time-dependent state onto $|0\rangle$. The FFT spectrum of the Rabi oscillations and the fitted transition amplitudes $|a_{i,1}|$ are plotted in Fig.~\ref{phasestatesweep}. 

Neglecting the counter-rotating terms (dashed lines), the two sidebands $\omega_m\pm (\lambda^+-\lambda^-)$ have the same amplitudes for both the phase  and initial state sweeps. 
However, experimentally, we measure a large asymmetry in the sideband amplitudes, with different asymmetry for the  phase sweep and the initial state sweep. 
Such an effect can be explained by the correction term $\frac{\epsilon_m^2}{8\Omega}\sigma_x$ in the Hamiltonian in Eq.~\eqref{HIeff} which makes the eigenstate shift from the y axis in the XY plane, introducing a dissimilarity between the phase and initial state sweeps. 
In the initial state sweep, a single mode evolution is achieved when the initial state is prepared along a shifted direction (see Fig.~\ref{phasestatesweep_Appendix}(d) in Appendix).
Tuning only the phase cannot instead achieve such a single mode evolution. Experiments show that the amplitudes of the sidebands deviate from the RWA prediction throughout all the ranges in Figs.~\ref{phasestatesweep}(c) and \ref{phasestatesweep}(d). 
The amplitude of the center band only has a large deviation from the RWA prediction when the initial state is close to the driving direction y in the ZY sweep in (d) with $\theta=\pi/2$. When the initial state is far away from the y direction, the Floquet prediction overlaps with the RWA prediction. 
The initial state sweep experiments in Fig.~\ref{phasestatesweep} and in Appendix Fig.~\ref{phasestatesweep_Appendix} show that for the center band,  $\delta|a_{0,1}|_{x}>\delta|a_{0,1}|_{y}>\delta|a_{0,1}|_{z}$, as predicted from the analysis, where $x,y,z$ denote the corresponding directions of the initial state. Whereas analysis with the effective Hamiltonian in Eq.~\eqref{HIeff} only provides a qualitative estimate, Floquet simulation accurately predicts the values of the amplitudes quantitatively for all experimental conditions.

In addition to frequency and amplitude shifts, we see frequency components $2\omega_m, 2\omega_m\pm(\lambda^+-\lambda^-)$ and even $3\omega_m$ appearing in the range of $f=7\sim 12\text{MHz}$. Even though their intensities are weak, these frequency components represent results that are beyond those predicted by the RWA. 

Thus, large deviation from the RWA predictions happens in the evolution mode control experiments, indicating drastic modifications caused by the counter-rotating terms. 
One application of the CCD scheme is to protect quantum information by tuning the evolution mode to the robust center band when the spin-locking condition is satisfied. Our experiment shows that such a condition is shifted and a simple phase adjustment is not enough to find the optimal condition when the driving strength is comparable with the static splitting. One solution is to tune the driving strength $\Omega$ to compensate for the Hamiltonian correction, as discussed in the following subsection.

\subsection{Resonance shifts}
\begin{figure}[t]
\includegraphics[width=86mm]{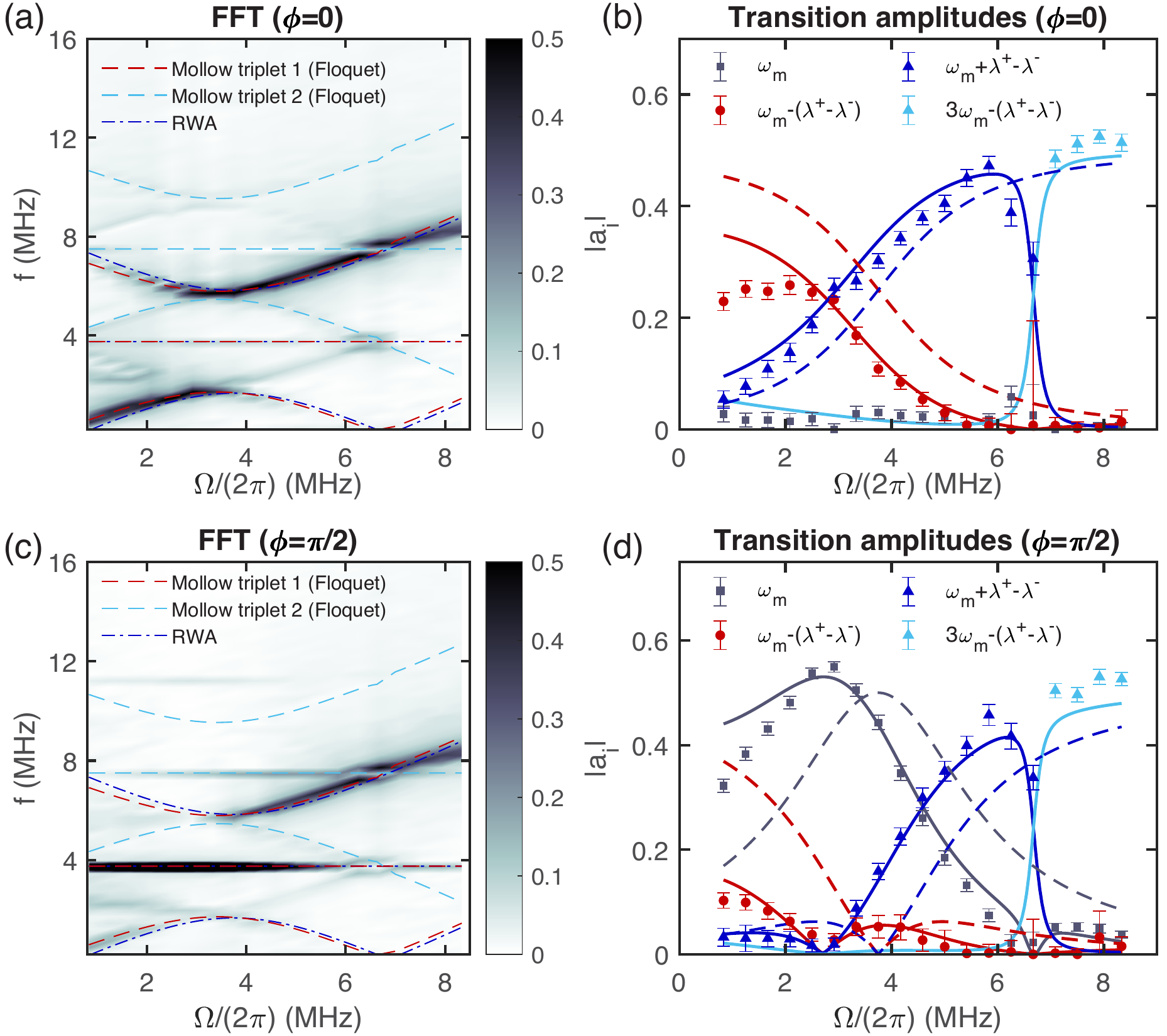}
\caption{
\label{freqcomponents_AvoidedCross}  
\textbf{Resonance shifts.} (a) $\Omega$-dependence of the Mollow triplet with parameters $\delta=0$, $\omega_m=(2\pi)3.75\text{MHz}$, $\epsilon_m = (2\pi)2.08\text{MHz}$, $\phi=0$. (b) Transition amplitudes $|a_{i,n}|$ extracted from the data in (a) (see Fig.~\ref{phasestatesweep}). The gray squares, red circles, blue triangles, and light blue triangles are experimental transition amplitudes $|a_{i,n}|$ of frequency components $\omega_m, \omega_m-(\lambda^+-\lambda^-), \omega_m+\lambda^+-\lambda^-,3\omega_m-(\lambda^+-\lambda^-)$, respectively. Solid lines and dashed lines correspond to the Floquet prediction and the RWA prediction. (c-d) Same as as in (a) except that the driving phase was set to $\phi=\pi/2$. 
Note that the data points in (b) and (d) share the same normalization factor to best match the Floquet prediction curves. Due to the power saturation when $\Omega/(2\pi)=6\sim 8\text{MHz}$, we add an additional $\epsilon\cos(2\omega_m t)\sigma_y$ term with $\epsilon=(2\pi)0.2\text{MHz}$ in the Floquet calculations  (Eq.~\eqref{HI_AmpMod}), as explained in Appendix \ref{app:addresults}.}
\end{figure}
Without the counter-rotating effect, in the first rotating frame (see Eq.~\eqref{HI_AmpMod}), the resonance condition is satisfied when $\omega_m=\Omega$ ($\delta=0$). We study the resonance shift caused by counter-rotating effects by experimentally sweeping the main driving strength $\Omega$ under two different modulation phases $\phi=0,\phi=\pi/2$ (see Figs.~\ref{freqcomponents_AvoidedCross}(a) and \ref{freqcomponents_AvoidedCross}(c)). 
The frequency shifts due to counter-rotating terms are very small when $\Omega>\epsilon_m$. 
A larger shift appears when $\Omega<(2\pi)2\text{MHz}$, and the Floquet prediction plotted with dashed red and light blue lines in Figs.~\ref{freqcomponents_AvoidedCross}(a) and \ref{freqcomponents_AvoidedCross}(c) showcases an improved fit on the experimental data. Figures~\ref{freqcomponents_AvoidedCross}(b) and \ref{freqcomponents_AvoidedCross}(d) are the oscillation amplitudes $|a_{i,n}|$ of the corresponding frequency components obtained by fitting the Rabi oscillation data. 
The Floquet predictions are plotted in solid lines while the RWA predictions are plotted in dashed lines.
When $\phi=\pi/2$, the disappearance of the sidebands happens at $\Omega\approx(2\pi)2.9\text{MHz}$ in (d), which is clearly on the left of the resonant frequency $(2\pi)3.75\text{MHz}$, indicating a change of the resonance condition. 
Similarly, when $\phi=0$, the crossing point of the two sidebands is also at $\Omega\approx(2\pi)2.9\text{MHz}$ in (b). 
In Eq.~\eqref{HIeff}, the detuning term $\frac{\Omega-\omega_m}{2}$ can be seen as a compensation of the correction $\frac{\epsilon_m^2}{8\Omega}$. 
A simple calculation predicts the compensation at $\Omega\approx(2\pi)3.4\text{MHz}$, which is still larger than the measured and simulated values of $\sim(2\pi)2.9\text{MHz}$, indicating that just including the first order correction in the effective Hamiltonian in Eq.~\eqref{HIeff} is not enough, and more corrections need to be taken into account. The Floquet calculation is an accurate way to predict such a shift. Note that when $\Omega$ is small, the measured amplitudes $|a_{i,n}|$ are lower than the  theoretical curve: this is due to the small Rabi contrast under weak driving (see details in Supplemental Material).  

In addition to the resonance shifts, we also observe higher-order frequency components corresponding to $n=2,3$ in the Mollow triplet. 
In Figs.~\ref{freqcomponents_AvoidedCross}(a) and \ref{freqcomponents_AvoidedCross}(c), we use the dashed red and light blue lines to plot the first (n=1) and second set (n=2) of frequency values of the Mollow triplet predicted by the Floquet theory.
In Fig.~\ref{freqcomponents_AvoidedCross}(c), we clearly observe the $2\omega_m$ and $3\omega_m$ components when $\Omega$ is small. In Fig.~\ref{freqcomponents_AvoidedCross}(a), we see the sidebands of the $n=3$ set around $\Omega/(2\pi)=1\text{MHz},f=8\text{MHz}$ and the sidebands with $n=2$ also exist but are partly hidden under the dashed lines.

\subsection{Frequency shifts and avoided crossing}
\begin{figure}[htbp]
\includegraphics[width=86mm]{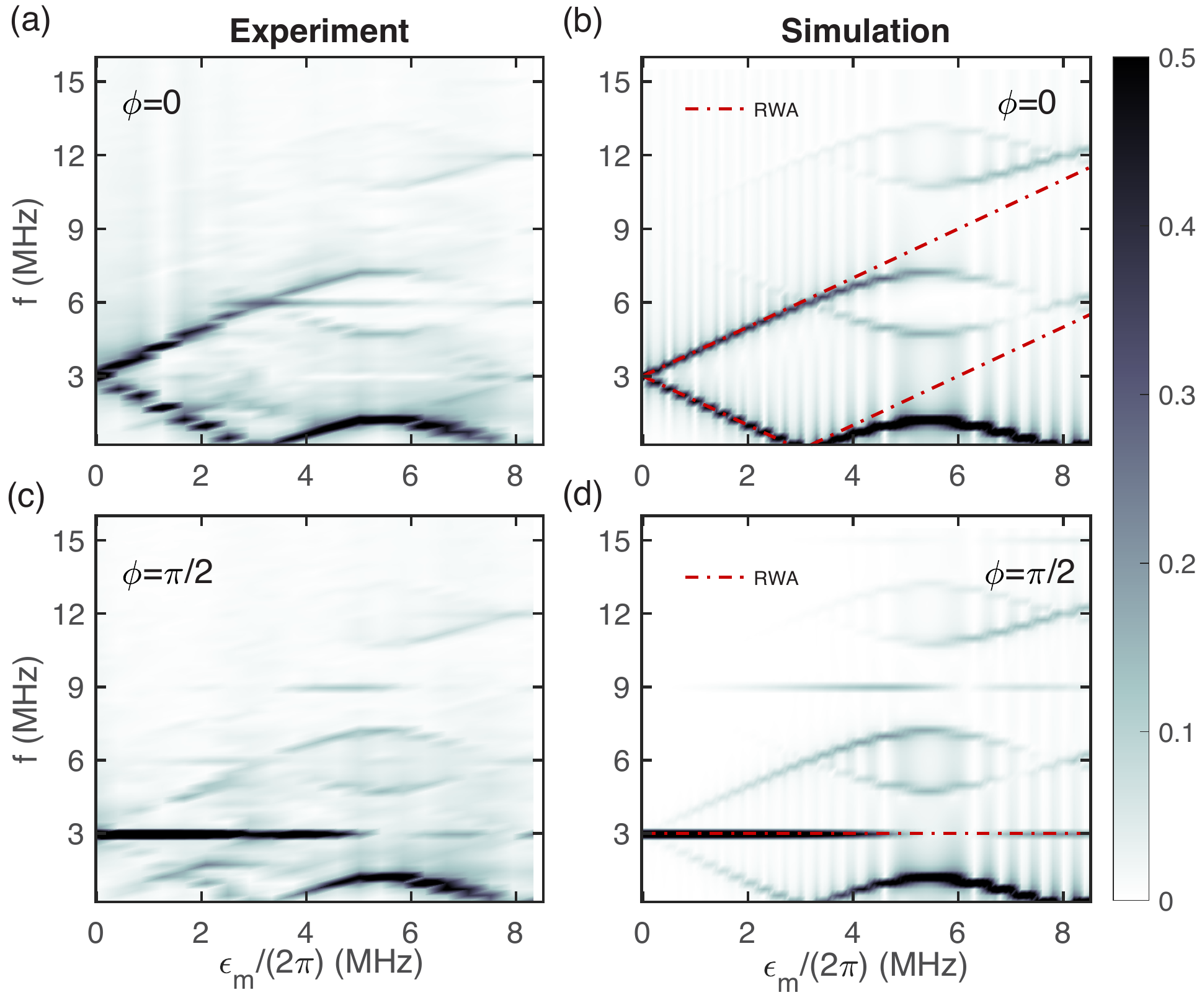}
\caption{\label{emsweep} \textbf{Energy shifts and avoided crossing in the Mollow triplet.} (a) and (c) are the Mollow triplet structure observed with the phase-modulated CCD at different values of $\phi$. (b) and (d) are  simulations by the Floquet approach with the same parameters. Parameters are $\delta=0,\Omega=\omega_m=(2\pi)3\text{MHz},\phi=0 \text{ in (a-b)},\phi=\pi/2\text{ in (c-d)}$. }
\end{figure}
Since the energy eigenvalues are less sensitive to the counter-rotating effects, a larger driving strength $\epsilon_m$ is needed to observe a frequency deviation from the RWA predictions. 
We swept the modulation strength $\epsilon_m$ from 0 to $\sim 3\Omega$ in the phase-modulated CCD scheme, on resonance and for two different driving field phases ($\phi=0,  \pi/2$).
The FFT spectra of the measured Rabi oscillations are plotted in Figs.~\ref{emsweep}(a) and \ref{emsweep}(c), while Figs.~\ref{emsweep}(b) and \ref{emsweep}(d) show the equivalent spectra simulated by the Floquet approach. We further compare the simulations to the predictions based on the  RWA (red dashed lines). 
When $\epsilon_m<(2\pi)2\text{MHz}$, three frequency components $\omega_m,\omega_m\pm\epsilon_m$ dominate the spectra. 
Within this region, both the Floquet and the RWA approaches correctly predict the frequencies of the three bands in the Mollow triplet structure, with the two sidebands linearly dependent on  $\epsilon_m$. 
When $\epsilon_m$ becomes larger, a pattern appears in the higher frequency region corresponding to the third manifold of the Mollow triplet structure $\omega_{i,3}$ centered at $3\omega_m=(2\pi)9\text{MHz}$, which follows a simple translation of the lowest order pattern centered at $\omega_m=3\text{MHz}$.
In our experiments, we can thus clearly observe at least two higher order Mollow triplets as predicted by Floquet simulations, in addition to the one predicted by the RWA. 
Note that the two sidebands frequencies   are no longer linearly dependent on $\epsilon_m$ and display a clear bending around $(2\pi)$4.8MHz  due to an avoided crossing. We show in Fig.~\ref{emsweep_freq} in Appendix that such avoided crossings are caused by the mixing between the frequency components $\omega_m-(\lambda^+-\lambda^-)$ and $\lambda^+-\lambda^-$ due to the counter-rotating terms. Such an avoided crossing only exists in the Floquet simulation. 

\section{Conclusion}
\label{sec:conclusion}
In this work, we explored higher order effects in the Mollow triplet structure by experimentally applying a modulated driving field (following the CCD scheme) and analyzing the results with Floquet theory. 
We observed frequency components beyond the RWA predictions,  shifts of the eigenenergies $\lambda^\pm$ and of the transition amplitudes  $|a_{i,n}|$. 
Our results not only demonstrate the ability of the modulated driving protocol to observe effects due to the strong driving, but also pave the way to employing this technique for robust control.
Indeed, as the $n\omega_m$  frequency components  of the Floquet evolution  are only determined by the modulation frequency $\omega_m$,  they are robust against external noise and fluctuations in the driving fields. Combined with our mode-control technique, this enables the generation of highly-robust quantum states \cite{wangCoherenceProtectionDecay2020}.  Nevertheless, to achieve such robustness, a precise knowledge of the Floquet eigenstates is needed to accurately select the initial state or the driving phase, as demonstrated by our experimental results. Thus,  insights  from Floquet theory can be used to design robust quantum operations and protect the quantum coherence \cite{wangCoherenceProtectionDecay2020}. 
In particular, our work shows that to achieve optimal quantum control with the CCD scheme, we should be careful of not only the eigenenergy shifts due to the counter-rotating terms, but also the amplitude changes and resonance shifts, both of which prove to be more sensitive. 
Beyond applications to robust quantum control, our results provide a versatile technique for studying the effects of strong driving, overcoming practical experimental limitations in reaching the strong-driving regime, that could be applied to investigate other phenomena such as coherent destruction of tunneling and dynamic localization. 

\section*{Acknowledgement}
This work was supported in part by DARPA DRINQS and NSF PHY1915218. We thank Pai Peng for fruitful discussions and Thanh Nguyen for manuscript revision.

\appendix
\section{Floquet theory}
\label{Derivation: Floquet}
Floquet theory can be used to solve the quantum dynamics under a Hamiltonian periodic in time \cite{shirleySolutionSchrodingerEquation1965}. Similar to Bloch theory which can solve a Hamiltonian periodic in space and gives rise to a series of band structures in k-space,  Floquet theory  also predicts a series of 'band' structures in frequency space. Although the principles of Floquet theory is simple and straightforward, the physics residing in its energy structure equips it with the ability to solve detailed dynamics of higher order phenomena such as multi-photon process \cite{chuFloquetTheoremGeneralized2004}. 
Given a time-periodic Hamiltonian $H(t)=H(t+\frac{2\pi}{\omega})$, the  wavefunction has the form $\Psi(t)=e^{-i\lambda t}\Phi(t)$ where $\Phi(t)=\Phi(t+\frac{2\pi}{\omega})$ is periodic in time and $\lambda$ denotes the eigenenergy of the system. 
To solve for the wavefunction time evolution, we can apply the following steps. 
(1) Decompose the Hamiltonian $H(t)$ and the state vector $\Phi(t)$ into a Fourier series $\sum_{n}H_n e^{-in\omega t}$ and $\Phi(t)=\Phi_n e^{-in\omega t}$.  (2) Write out the Floquet matrix $H_F$ (see Eq.~\eqref{eq:FloquetMatrix}).
(3) Solve for the eigenvalue problem of the Floquet matrix $H_F\Phi=\lambda\Phi$. (4) Apply the initial conditions to get the  coefficients of each eigenstate $\Psi(t=0)=\sum_\alpha c^\alpha \sum_{n=-\infty}^{+\infty}\Phi_n^\alpha$. (5) Obtain the evolution $\Psi(t)=\sum_\alpha c^\alpha \Psi^\alpha(t)$.
Below we explicitly show how to implement this procedure.

\subsection{\label{sec:level1} General derivation of Floquet theory}
To solve the Schr\"{o}dinger equation $i\frac{\partial}{\partial t}\Psi(t)=H(t)\Psi(t)$, we obtain the eigenvalue problem for the periodic part $\Phi(t)$,
\begin{equation}
    \left(H(t)-i\frac{\partial}{\partial t}\right)\Phi(t)=\lambda\Phi(t)
\end{equation}
Plugging in the Fourier expansions $\Phi(t)=\Phi_n e^{-in\omega t},H(t)=\sum_{n}H_n e^{-in\omega t}$, we obtain
\begin{align}
\left(\sum_{n}H_n(t)e^{-in\omega t}-i\frac{\partial}{\partial t}\right)\sum_m\Phi_me^{-im\omega t}=\lambda \sum_m\Phi_me^{-im\omega t}
\end{align}
Writing out equation above in  matrix form, we obtain
\begin{equation} 
\begin{pmatrix}
 \ddots&\vdots &0&\vdots& \\ 
\cdots &H_0\!+\!\omega&H_{-1}&0&\cdots \\ 
 \cdots&H_1&H_0&H_{-1}& \\ 
 \cdots&0&H_1&H_0\!-\!\omega &\cdots \\ 
 &\vdots&\vdots&\vdots&\ddots
\end{pmatrix}\!\!
\begin{bmatrix}
\vdots\\
\Phi_{-1}\\
\Phi_0\\
\Phi_1\\
\vdots
\end{bmatrix}\!\!
=\!\lambda\!
 \begin{bmatrix}
\vdots\\
\Phi_{-1}\\
\Phi_0\\
\Phi_1\\
\vdots
\end{bmatrix}
\label{eq:FloquetMatrix}
\end{equation}
where the matrix in Eq.~\eqref{eq:FloquetMatrix}  is the Floquet matrix $H_F$. 

By solving for the eigenenergies $\lambda^\alpha$ and  eigenvectors $(\cdots,\Phi_{-1}^\alpha,\Phi_0^\alpha,\Phi_1^\alpha,\cdots)^T$ of the Floquet matrix, one can find the time-dependent energy eigenvectors of the system $\Psi^\alpha(t)=\sum_{n=-\infty}^{+\infty}e^{-i\lambda^\alpha t-in\omega t}\Phi_n^\alpha
    =e^{-i\lambda^\alpha t}\sum_{n=-\infty}^{+\infty}e^{-in\omega t}\Phi_n^\alpha$.
The evolution of the system can then be expressed as a superposition of these eigenvectors
\begin{equation}
    \Psi(t)=\sum_\alpha c^\alpha \Psi^\alpha(t)= \sum_\alpha c^\alpha e^{-i\lambda^\alpha t}\sum_{n=-\infty}^{+\infty}e^{-in\omega t}\Phi_n^\alpha
\end{equation}
with coefficients $c^\alpha$ determined by the initial condition
\begin{equation}
    \Psi(t=0)=\sum_\alpha c^\alpha \sum_{n=-\infty}^{+\infty}\Phi_n^\alpha.
\end{equation}. 

\paragraph{Quasi energy}
If $\lambda^\alpha$ is the eigenenergy of the system, $\lambda^\alpha+n\omega$ for any integer n is also the eigenenergy of the system since $\Psi^\alpha(t)=e^{-i\lambda^\alpha t}\Phi^\alpha(t)=e^{-i\lambda^\alpha t-in\omega t}(e^{in\omega t}\Phi^\alpha(t))$. The eigenfunction $\Psi^\alpha$ for the eigenenergy $\lambda^\alpha$ is the same as the eigenfunction $\Psi^{\alpha,n\omega}$ for the eigenenergy $\lambda^\alpha+n\omega$. As a result, it is usually possible to limit the range of the eigenvalues within the first ``Brillouin zone''  $[0,\omega)$ since all eigenvalues in the other zones will be a simple (frequency) translation of the values in the first zone. 
\paragraph{Number of solutions in a two level systems}
For two level systems, the Hamiltonian $H(t)$ can be written as a $2\times2$ matrix. The wavefunction $\Psi(t)$, its periodic part $\Phi(t)$, and the Fourier components $\Phi_n$,  are two dimensional vectors. Thus,  there are only two non-trivial solutions denoted by $\lambda^\pm$.

\subsection{Concatenated continuous driving}
We consider the Hamiltonian of the amplitude-modulated concatenated continuous driving in the interaction picture $H_I=-\frac{\delta}{2}\sigma_z+\frac{\Omega}{2}\sigma_x+\epsilon_m\cos(\omega_mt+\phi)\sigma_y$. Fourier decomposition of Hamiltonian gives $H_{I,0} = -\frac{\delta}{2}\sigma_z+\frac{\Omega}{2}\sigma_x,\ H_{I,\pm1} = \frac{\epsilon_m e^{\pm i\phi}}{2}\sigma_y$. Applying the Floquet approach to solve for the exact evolution of the system we obtain the following eigenvalue equation
\begin{widetext}
\begin{equation} 
\begin{pmatrix}
 \ddots&\vdots&\vdots&\vdots&\vdots&\vdots&\vdots& \\ 
\cdots &-\frac{\delta}{2}+\omega_m &\frac{\Omega}{2} &0 &-\frac{\epsilon_m}{2}ie^{i\phi} &0&0&\cdots \\ 
 \cdots&\frac{\Omega}{2} &\frac{\delta}{2}+\omega_m &\frac{\epsilon_m}{2}ie^{i\phi} &0 &0 &0 &\cdots\\
 \cdots&0 &-\frac{\epsilon_m}{2}ie^{-i\phi} & -\frac{\delta}{2} &\frac{\Omega}{2}&0&-\frac{\epsilon_m}{2}ie^{i\phi}&\cdots\\
 \cdots&\frac{\epsilon_m}{2}ie^{-i\phi} &0 &\frac{\Omega}{2} &\frac{\delta}{2} &\frac{\epsilon_m}{2}ie^{i\phi}&0 &\cdots\\
 \cdots&0 &0 &0 &-\frac{\epsilon_m}{2}ie^{-i\phi}&-\frac{\delta}{2}-\omega_m &\frac{\Omega}{2} &\cdots\\
 \cdots&0 &0 &\frac{\epsilon_m}{2}ie^{-i\phi} &0 &\frac{\Omega}{2} &\frac{\delta}{2}-\omega_m &\cdots\\
 &\vdots&\vdots&\vdots&\vdots&\vdots&\vdots&\ddots
\end{pmatrix}
\begin{pmatrix}
\vdots\\
\Phi_{-1,0}\\
\Phi_{-1,1}\\
\Phi_{0,0}\\
\Phi_{0,1}\\
\Phi_{1,0}\\
\Phi_{1,1}\\
\vdots
\end{pmatrix}
=\lambda
\begin{pmatrix}
\vdots\\
\Phi_{-1,0}\\
\Phi_{-1,1}\\
\Phi_{0,0}\\
\Phi_{0,1}\\
\Phi_{1,0}\\
\Phi_{1,1}\\
\vdots
\end{pmatrix}
\end{equation}
The evolution of the system can be written in terms of the Floquet energy eigenvalues and eigenvectors, taking into account the initial conditions:
	\begin{equation}
	\Psi(t)=c^+e^{-i\lambda^+t}\sum_n 
\begin{pmatrix}
    \Phi^+_{n,0}\\
    \Phi^+_{n,1}
    \end{pmatrix}
e^{-in\omega t}    +c^-e^{-i\lambda^-t}\sum_n 
\begin{pmatrix}
    \Phi^-_{n,0}\\
    \Phi^-_{n,1}
    \end{pmatrix}
e^{-in\omega t}=    \Psi(t)=\sum_n 
\begin{pmatrix}
    c^+e^{-i\lambda^+t}\Phi^+_{n,0}+c^-e^{-i\lambda^-t}\Phi^-_{n,0}\\
    c^+e^{-i\lambda^+t}\Phi^+_{n,1}+c^-e^{-i\lambda^-t}\Phi^+_{n,1}
    \end{pmatrix}e^{-in\omega t}	
	\end{equation}    
%
The system evolution is probed experimentally by measuring the system population  by projecting the state onto $|0\rangle$. The probability of being in the $|0\rangle$ state $P_{|0\rangle}(t)$ presents three classes of frequencies: $n\omega\pm(\lambda^+-\lambda^-)$ and $n\omega$, with $n$ integer:
%
\begin{align}
        P_{|0\rangle}(t)&=\sum_n e^{-in\omega t}(c^+e^{-i\lambda^+t}\Phi^+_{n,0}+c^-e^{-i\lambda^-t}\Phi^-_{n,0})\times
        \sum_m e^{im\omega t}(c^{+*}e^{i\lambda^+t}\Phi^{+*}_{m,0}+c^{-*}e^{i\lambda^-t}\Phi^{-*}_{m,0})\\
        &=\sum_{n,m}e^{i(m-n)\omega t}\left(|c^+|^2\Phi_{m,0}^{+*}\Phi_{n,0}^{+}+|c^-|^2\Phi_{m,0}^{-*}\Phi_{n,0}^{-}+
        e^{-i(\lambda^+-\lambda^-)t}c^+c^{-*} \Phi_{n,0}^{+}\Phi_{m,0}^{-*}
        +e^{i(\lambda^+-\lambda^-)t}c^{+*}c^{-}\Phi_{m,0}^{+*}\Phi_{n,0}^{-}\right)\nonumber
\end{align}
\end{widetext}
We can rewrite this expression as 
\begin{equation}
  P_{|0\rangle}(t)  = \sum_{i,n} |a_{i,n}|\cos(\omega_{i,n} t+\phi_{i,n})
\end{equation}        
with $\omega_{i,n}=n\omega_m+i(\lambda^+-\lambda^-)$, where $i=-1,0,1$, $a_{\pm 1,n} = |a_{\pm 1,n}|\exp(i\phi_{\pm 1,n})=2\sum_k c^{\pm*}c^{\mp}\Phi_{k+n,0}^{\pm*}\Phi_{k,0}^{\mp}$, and $a_{0,n}=2\sum_{\pm}|c^\pm|^2\sum_k \Phi_{k+n,0}^{\pm*}\Phi_{k,0}^{\pm}$
  The first few frequencies and coefficients are listed in Table \ref{P0_appendix}.
\begin{table}[htbp]
\caption{Components of time dependent population $P_{|0\rangle}(t)$.}
\label{P0_appendix}
\begin{tabular}{c|c}
\hline\hline
Frequency $\omega_{i,n}$ & Coefficients $a_{i,n}$\\\hline
\hline
0 & $\sum_{\pm,k}|c^\pm|^2|\Phi^\pm_{k,0}|^2$ \\
\hline
$\lambda^+-\lambda^-$ & $2\sum_{k}c^{+*}c^{-}\Phi_{k,0}^{+*}\Phi_{k,0}^{-}$\\
\hline
$\omega_m$ & $2\sum_{\pm}|c^\pm|^2\sum_k \Phi_{k+1,0}^{\pm*}\Phi_{k,0}^{\pm}$ \\
\hline
$\omega_m-(\lambda^+-\lambda^-)$ & $2\sum_k c^{+}c^{-*}\Phi_{k,0}^{+}\Phi_{k+1,0}^{-*}$ \\
\hline
$\omega_m+(\lambda^+-\lambda^-)$ & $2\sum_k c^{+*}c^{-}\Phi_{k+1,0}^{+*}\Phi_{k,0}^{-}$ \\
\hline
$\cdots$ & $\cdots$ \\
\hline\hline
\end{tabular}
\end{table}

When the RWA  is valid, $\epsilon_m\ll\omega_m,\Omega$, and the resonant condition $\omega_m\approx\Omega$ is satisfied, counter-rotating terms can be dropped yielding a block-diagonal  Floquet Hamiltonian. (Note that a frame transformation that aligns the static field along z and the oscillating field along x is needed to obtain this block-diagonal form.) 
Following such procedure gives $\lambda^+-\lambda^- = \sqrt{\epsilon_m^2+(\omega_m-\Omega_R)^2}$, where $\Omega_R=\sqrt{\Omega^2+\delta^2}$ and the only non-vanishing oscillating components in the population are the first five components listed in Table \ref{P0_appendix}. When the approximation  is not valid, more frequency components are involved and the evolution is made more complicated.

\subsection{Two definitions of $\lambda^\pm$}
\begin{figure}[h]
\includegraphics[width=86mm]{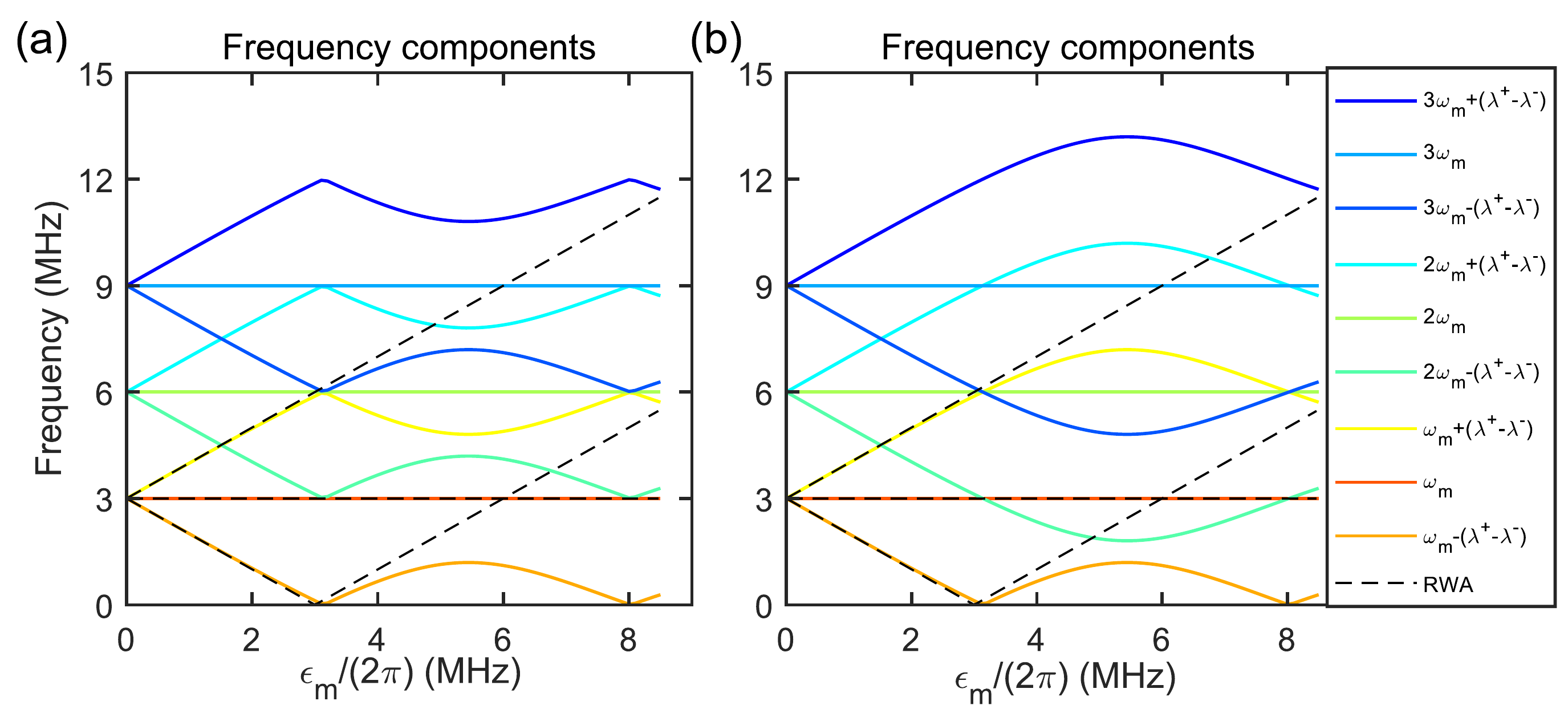}
\caption{\label{emsweep_freq}  
\textbf{Frequency components predicted by Floquet theory.}  Frequency components predicted by Floquet theory corresponding to the measured frequency components shown in Fig.~\ref{emsweep}. There are two ways to represent the frequencies: in (a) $\lambda^+$, $\lambda^-$ are limited to the first zone $[0,\omega_m)$ where $\omega_m=(2\pi)3\text{MHz}$. In (b) $\lambda^+$, $\lambda^-$ change smoothly as we increase  $\epsilon_m$.}
\end{figure}

As discussed above, there are two nontrivial solutions $\lambda^+,\lambda^-$ for the Floquet matrix up to a translation of $n\omega_m$ where n is any integer. One definition of $\lambda^\pm$ is by limiting their values within the first zone $[0,\omega_m)$, which is used in the numerical simulations in this work to simplify the calculation. An example of such a definition corresponding to Fig.~\ref{emsweep} is shown in Fig.~\ref{emsweep_freq}(a). In this paper, we use this definition for the amplitude calculation in Figs.~\ref{freqcomponents}(b) and \ref{freqcomponents}(d) and Figs.~\ref{freqcomponents_AvoidedCross}(b) and \ref{freqcomponents_AvoidedCross}(d) where we see a sudden switch between the frequency components $\omega_m+\lambda^+-\lambda^-$ and $3\omega_m-(\lambda^+-\lambda^-)$ when $\Omega\approx (2\pi)6.8\text{MHz}$. However, this definition makes it difficult to clarify whether we observe frequency components beyond the RWA. To clarify that we observe higher order frequency components $\omega_{i,n}$ with $n>1$, we use a definition that has a correspondence to the prediction obtained from the analytical RWA approach and no longer limits the range of $\lambda^\pm$. Figure~\ref{emsweep_freq}(b) shows such a definition where each manifold of Mollow triplet is clearly seen. In Figs.~\ref{freqcomponents}(a) and \ref{freqcomponents}(c) and Figs.~\ref{freqcomponents_AvoidedCross}(a) and \ref{freqcomponents_AvoidedCross}(c), the dashed lines are the frequency prediction from the Floquet theory using the same definition.

\section{Mode control of the evolution and resonance shifts}\label{app:addresults}
\begin{figure}[h]
\includegraphics[width=86mm]{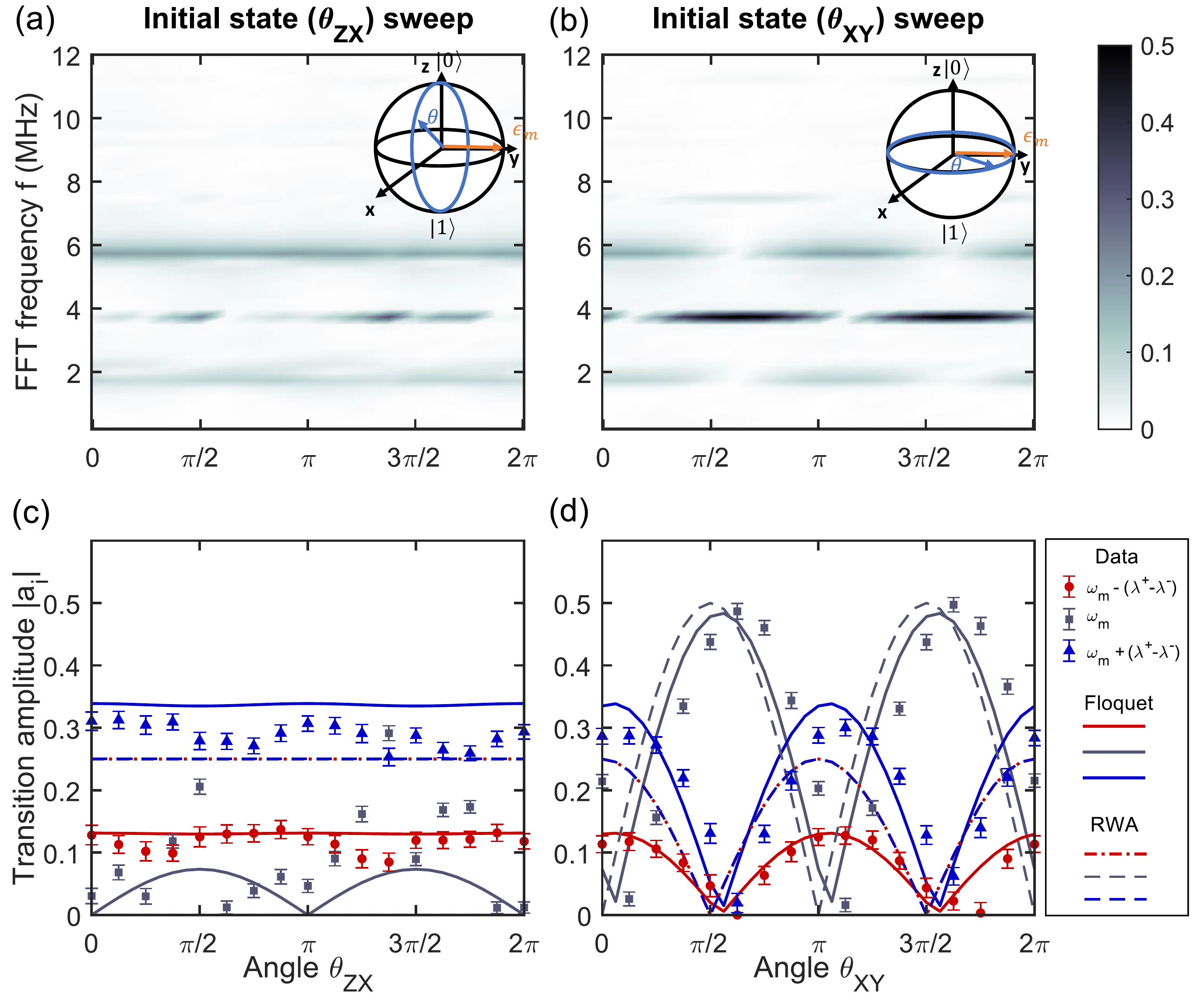}
\caption{\label{phasestatesweep_Appendix} 
\textbf{Evolution mode control and sideband asymmetry for state-swept experiments.}  (a) and (c) are FFT spectra and transition amplitude $|a_i|$ of the initial state sweep in ZX plane.  (b) and (d) are FFT and $|a_i|$ plots of the initial state sweep in XY plane. Parameters are $\Omega=\omega_m=(2\pi)3.75\text{MHz},\epsilon_m=(2\pi)2.08\text{MHz},\phi=0$.}
\end{figure}

In the main text we demonstrate control of the system evolution by sweeping the driving phase and the initial state. This enables tuning the evolution to highlight different frequency modes. Here we provide additional demonstration of such mode control by sweeping the initial state 
in the ZX plane and XY plane (see Fig.~\ref{phasestatesweep_Appendix}). These results further highlight the need to take counter-rotating effects when describing the system evolution under strong driving.

\begin{figure}[h]
\includegraphics[width=86mm]{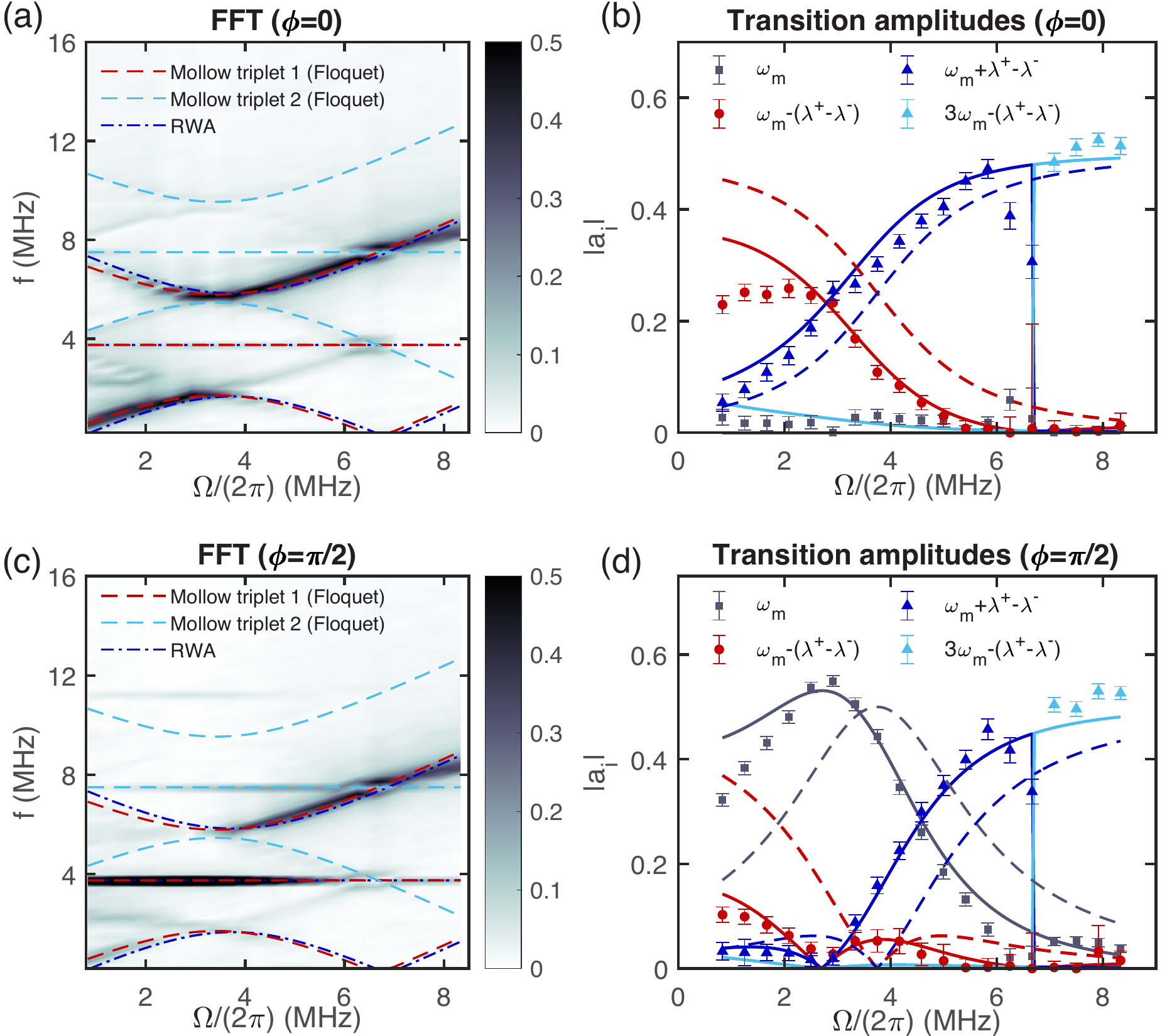}
\caption{\label{freqcomponents} \textbf{ Resonance shifts. }(a) $\Omega$-dependence of Rabi spectra with parameters $\delta=0,\omega_m=(2\pi)3.75\text{MHz},\epsilon_m = (2\pi)2.08\text{MHz}, \phi=0$. (b) Transition amplitude $|a_{i,n}|$ for frequency component in (a). Symbols are experimental data, solid lines the Floquet  predictions while dashed lines are RWA predictions.  (c) $\Omega$-dependence of Rabi  spectra with  parameters $\delta=0,\omega_m=(2\pi)3.75\text{MHz},\epsilon_m = (2\pi)2.08\text{MHz}, \phi=\pi/2$. (d) Transition amplitude $|a_{i,n}|$ for frequency component in (c) following the same conventions as in (b). Note that the data points in (b) and (d) share the same normalization factor to best match the theoretical prediction curves.}
\end{figure}

As we increase the driving strength to observe these effects we reach the saturation limits of our apparatus. This leads to unwanted, albeit interesting, additional modulations of the dynamics. 

We reproduce in Fig.~\ref{freqcomponents} the results shown in the main text (Fig.~\ref{freqcomponents_AvoidedCross}) but in the Floquet simulations we do not take into account saturation effects due to  imperfect electronics and power saturation of the amplifier. When we 
add an additional term $\epsilon\cos(2\omega_m t)$ to the Hamiltonian in Eq.~\eqref{HI_AmpMod} to mimic the saturation effects, we reproduce the gradual switch between frequency components $3\omega_m-(\lambda^+-\lambda^-)$ at $\Omega\approx (2\pi)6.8\text{MHz}$. Instead  the Floquet simulation without  any additional terms predicts that the switching happens suddenly at  a single point (see Fig.~\ref{freqcomponents} and  Supplemental Material for more details on the power saturation and avoided crossings).

\newpage
\bibliography{ModulatedDrivingFloquet} 
\bibliographystyle{apsrev4-1}

\pagebreak
\widetext
\setcounter{section}{0}
\setcounter{equation}{0}
\setcounter{figure}{0}
\setcounter{table}{0}
\setcounter{page}{1}
\makeatletter
\renewcommand{\theequation}{S\arabic{equation}}
\renewcommand{\thefigure}{S\arabic{figure}}


\begin{CJK*}{UTF8}{}
\title{Supplemental Material}

\author{Guoqing Wang \CJKfamily{gbsn}(王国庆)}
\affiliation{
   Research Laboratory of Electronics and Department of Nuclear Science and Engineering, Massachusetts Institute of Technology, Cambridge, MA 02139, USA}
\author{Yi-Xiang Liu \CJKfamily{gbsn}(刘仪襄)}
\affiliation{
   Research Laboratory of Electronics and Department of Nuclear Science and Engineering, Massachusetts Institute of Technology, Cambridge, MA 02139, USA}
\author{Paola Cappellaro}\email[]{pcappell@mit.edu}
\affiliation{
   Research Laboratory of Electronics and Department of Nuclear Science and Engineering, Massachusetts Institute of Technology, Cambridge, MA 02139, USA}
\affiliation{Department of Physics, Massachusetts Institute of Technology, Cambridge, MA 02139, USA}

\maketitle
\end{CJK*}

\begin{figure*}[htbp]
\includegraphics[scale = 0.48]{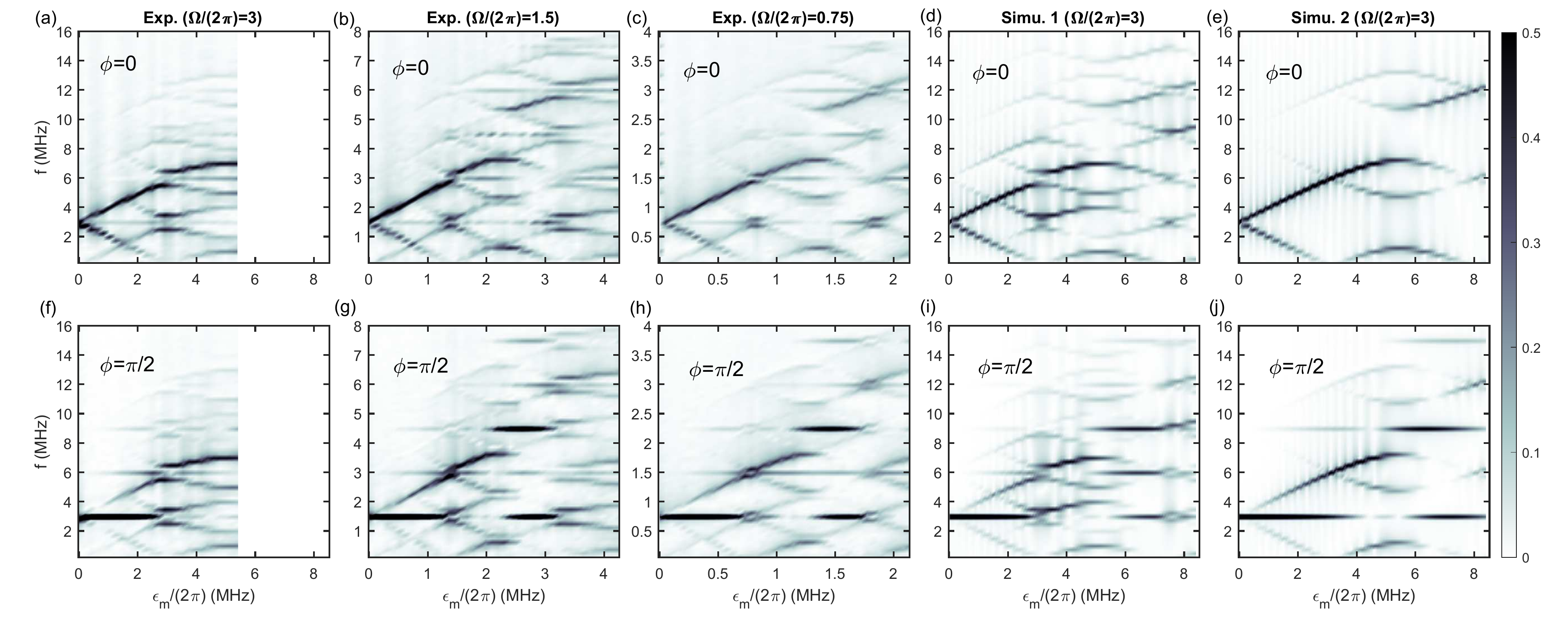}
\caption{\label{emsweep_dataSim}  \textbf{$\epsilon_m$ dependence of Rabi Fourier spectra}. (a) and (f) are amplitude-modulated experiments measured with $\Omega=\omega_m=(2\pi)3\text{MHz},\phi=0,\pi/2$ correspondingly. (b) and (g) are similar experiments measured with $\Omega=\omega_m=(2\pi)1.5\text{MHz}$, which scale as half of the parameters in (a) and (f). (c) and (h) are similar experiments with $\Omega=\omega_m=(2\pi)0.75\text{MHz}$, which scale as $1/4$ of the parameters in (a) and (f). (e) and (j) are simulations with Floquet theory using Hamiltonian $H_I=\frac{-\delta}{2}\sigma_z+\frac{\Omega}{2}\sigma_x+\epsilon_m\cos(\omega_m t+\phi).$ (d) and (i) are simulations with Floquet theory using Hamiltonian $H_I^{'} =H_I+\epsilon\cos(2\omega_m t)$ where $\epsilon=0.4\epsilon_m$ is the strength of the second order modulation.}
\end{figure*}

\begin{figure*}[h]
\includegraphics[scale = 0.625]{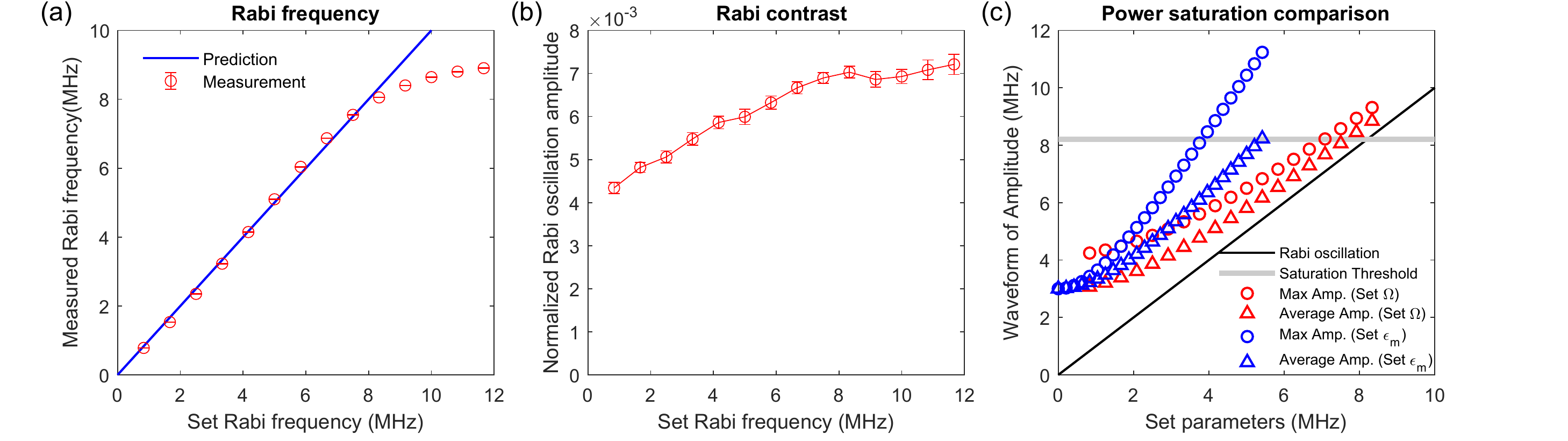}
\caption{\label{RabiFreqAmp_Saturation} (a) Rabi frequency versus voltage setting of the arbitrary waveform generator. The x axis is the nominal Rabi frequency as set by the AWG voltage setting  and the y axis is the measured Rabi frequency. Red points are the data and blue curve is linear prediction from the voltage setting. 
(b) Rabi oscillation amplitude under different Rabi frequencies. 
(c) Power saturation. For red triangles and circles, x axis is the driving strength of the microwave $\Omega/(2\pi)$ in the $\Omega/(2\pi)$ dependence experiments in the main text, y axis for the circles is the maximum amplitude of the waveform and for the triangles is the average amplitude of the waveform. For blue triangles and circles, x axis is the $\epsilon_m/(2\pi)$ in the $\epsilon_m$ sweep experiments in Fig.~\ref{emsweep_dataSim}(a) and Fig.~\ref{emsweep_dataSim}(f) and y axis is shared with the red points. The gray line is the saturation amplitude.}
\end{figure*}
\begin{figure*}[htbp]
\includegraphics[scale = 0.625]{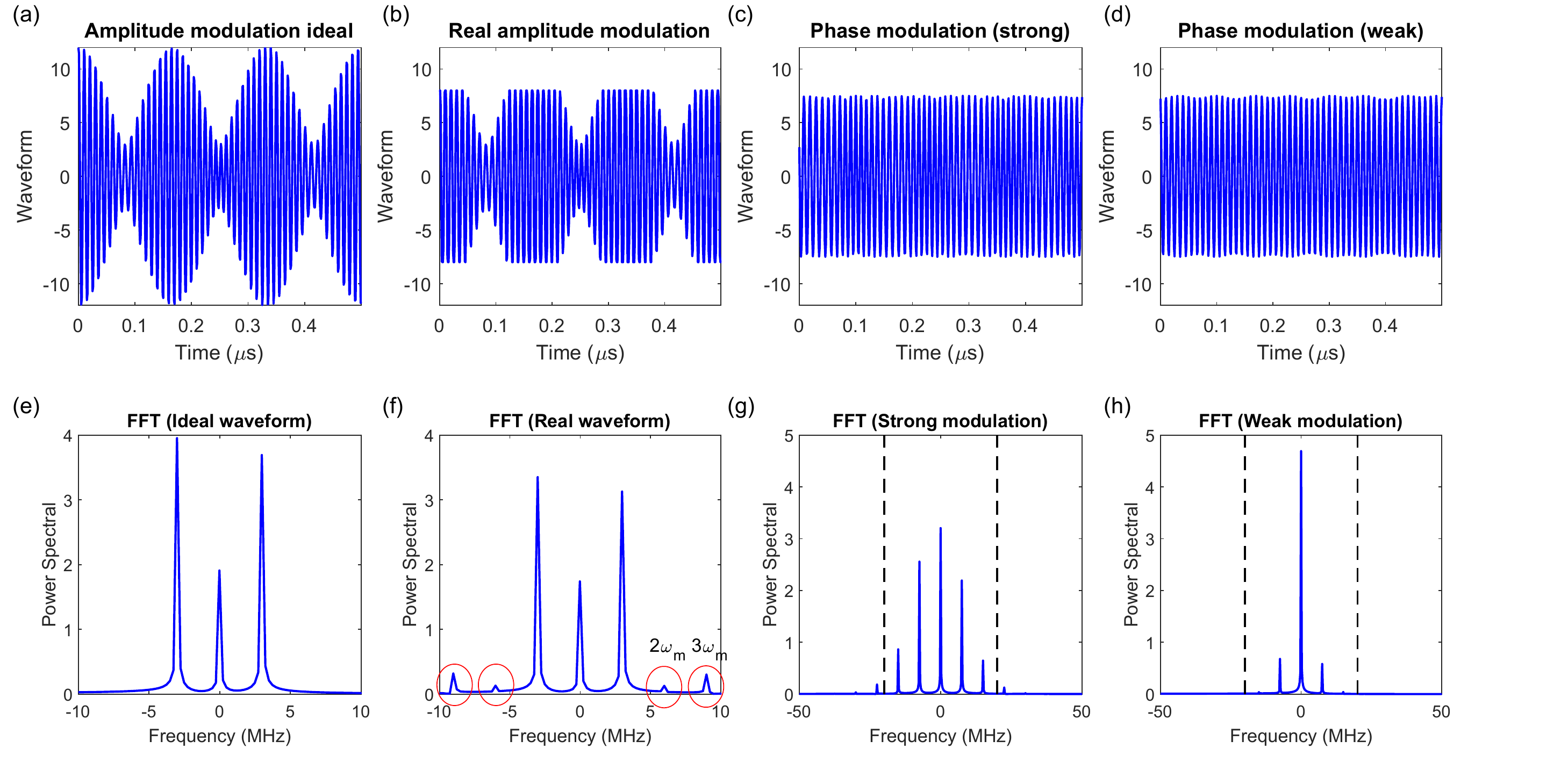}
\caption{
\label{RabiSaturationFFT} 
(a) Ideal waveform for the amplitude-modulated driving,  $\Omega\cos(\omega_0 t)-2\epsilon_m\sin(\omega_0 t)\cos(\omega_m t)$ with $\omega_0=(2\pi)100\text{MHz},\Omega=\omega_m=(2\pi)3\text{MHz},\epsilon=(2\pi)6\text{MHz}$. 
(b) Effective waveform of the driving term in (a) if we set a saturation level at $\pm 8$. Voltage exceeding this saturation threshold is set to the threshold value. (c) Phase-modulated waveform $\Omega\cos(\omega_0 t+2\epsilon_m/\Omega\cos(\omega_m t))$ with $\omega_0=(2\pi)100\text{MHz},\Omega=\omega_m=(2\pi)7.5\text{MHz},\epsilon=(2\pi)4.5\text{MHz}$. (d) Phase-modulated waveform $\Omega\cos(\omega_0 t+2\epsilon_m/\Omega\cos(\omega_m t))$ with $\omega_0=(2\pi)100\text{MHz},\Omega=\omega_m=(2\pi)7.5\text{MHz},\epsilon=(2\pi)1\text{MHz}$. (e-h) FFT spectra of the waveforms in (a-d) respectively. Note that frequency x axis   has been shifted by subtracting $\omega_0$ to prominently display the order of the newly appearing peaks.}
\end{figure*}

\begin{figure*}[htbp]
\includegraphics[scale = 0.8]{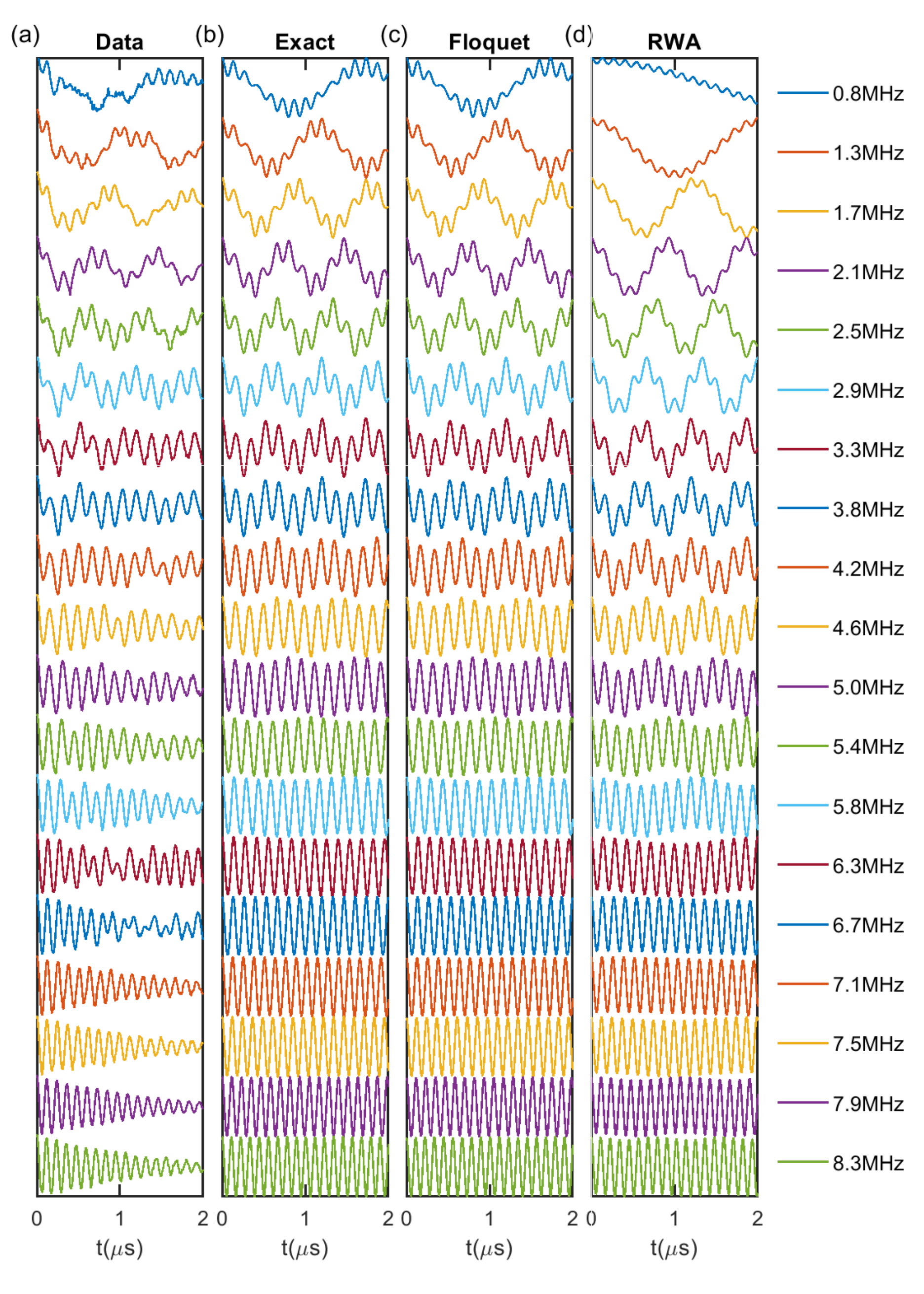}
\caption{\label{RabiCompare_Phi0} 
Raw data and simulations of Rabi oscillations in the $\Omega$ dependence experiments in main text with $\phi=0$. The legend is the value of $\Omega/(2\pi)$. (a) Rabi oscillation data. (b) Rabi oscillation calculated by directly evolving the Hamiltonian in a trotterized manner. (c) Floquet simulation by summing over the first 5 manifolds of triplet frequency components. (d) RWA simulation. Practically the RWA simulation is done in a same way as the Floquet simulation with the counter-rotating terms dropped in the Floquet matrix.}
\end{figure*}

\begin{figure*}[htbp]
\includegraphics[scale = 0.8]{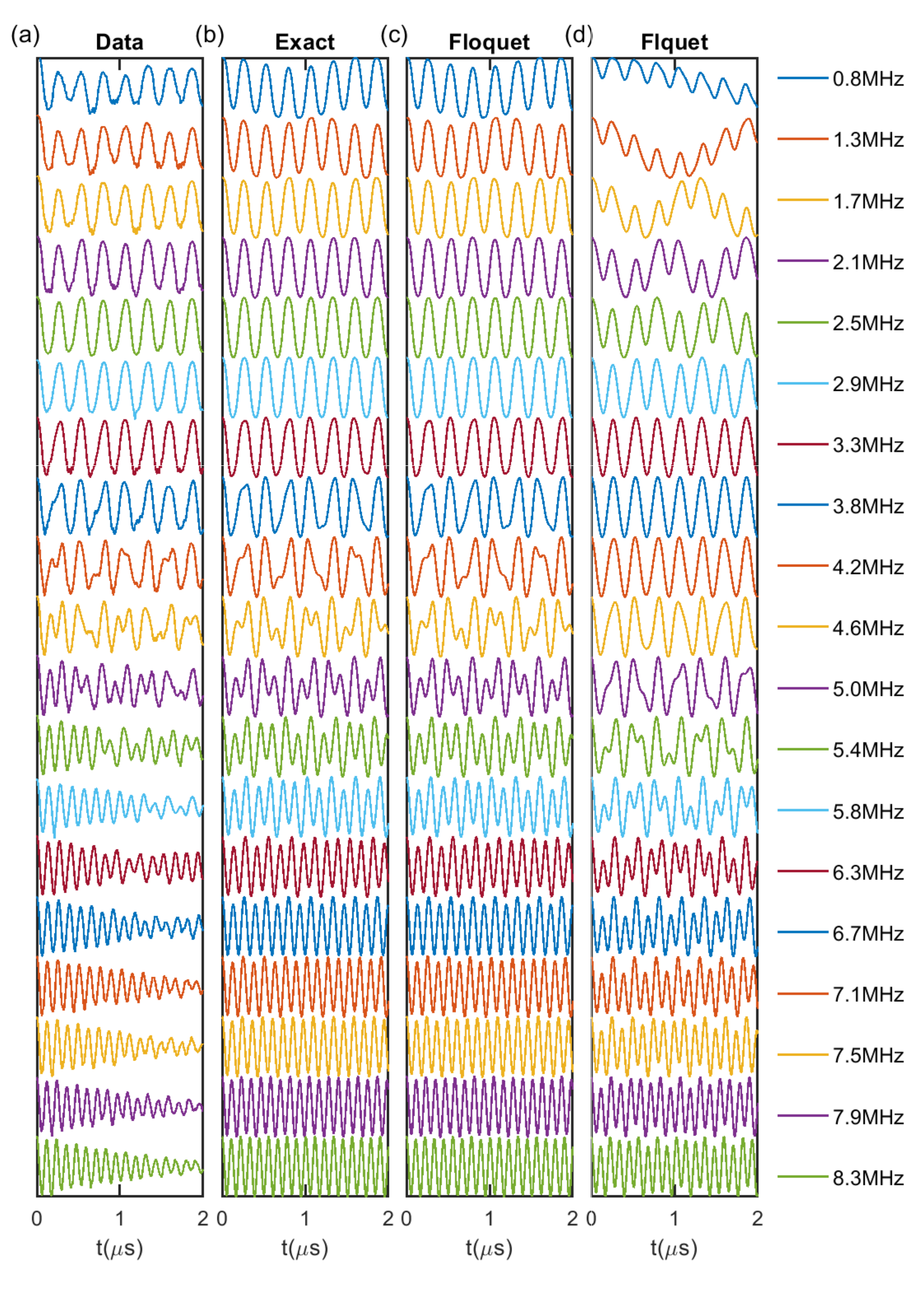}
\caption{\label{RabiCompare_PhiPio2} Raw data and simulations of Rabi oscillations in the $\Omega$ dependence experiments in main text with $\phi=\pi/2$. The legend is the value of $\Omega/(2\pi)$. (a) Rabi oscillation data. (b) Rabi oscillation calculated by direct evolving the Hamiltonian in a trotterized manner. (c) Floquet simulation by summing over the first 5 manifolds of triplet frequency components. (d) RWA simulation.}
\end{figure*}

\begin{figure*}[htbp]
\includegraphics[scale = 0.8]{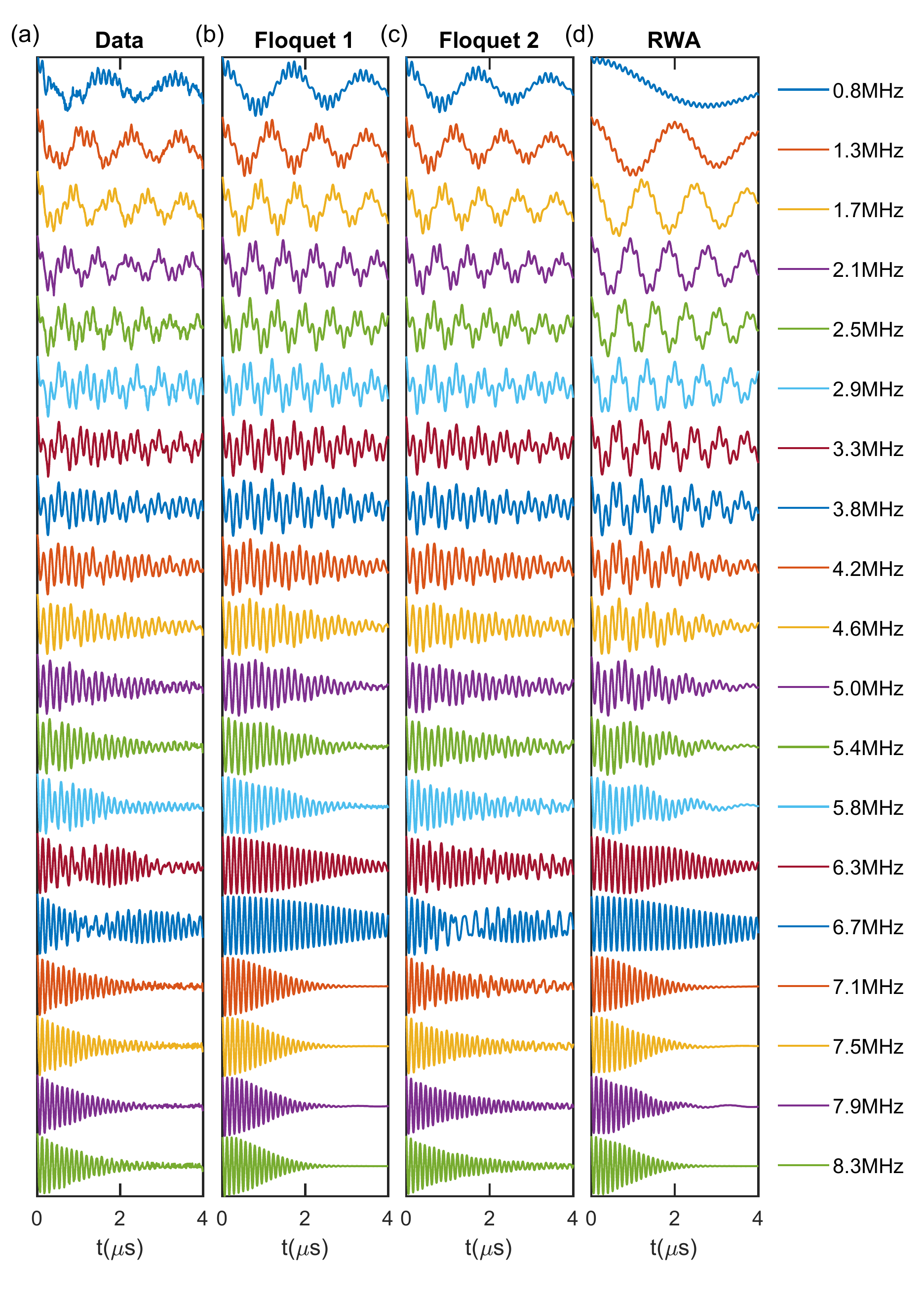}
\caption{\label{RabiCompare_Phi0_Decay} Raw data and simulations of Rabi oscillations in the $\Omega$ dependence experiments in main text with $\phi=0$. The legend is the value of $\Omega/(2\pi)$. (a) Rabi oscillation data. (b) Floquet simulation with decay. Decay factors $\exp(-t/\tau_{i,n})$ are multiplied to each frequency component with $\tau_{i,n}$ fitted from the experiment data. (c) Floquet simulation with decay and avoided crossing by adding an additional term $\epsilon\cos(2\omega_m t)\sigma_y$ to the Hamiltonian in Eq. (1) of the main text and implementing the Floquet simulation. (d) RWA simulation. }
\end{figure*}

\begin{figure*}[htbp]
\includegraphics[scale = 0.8]{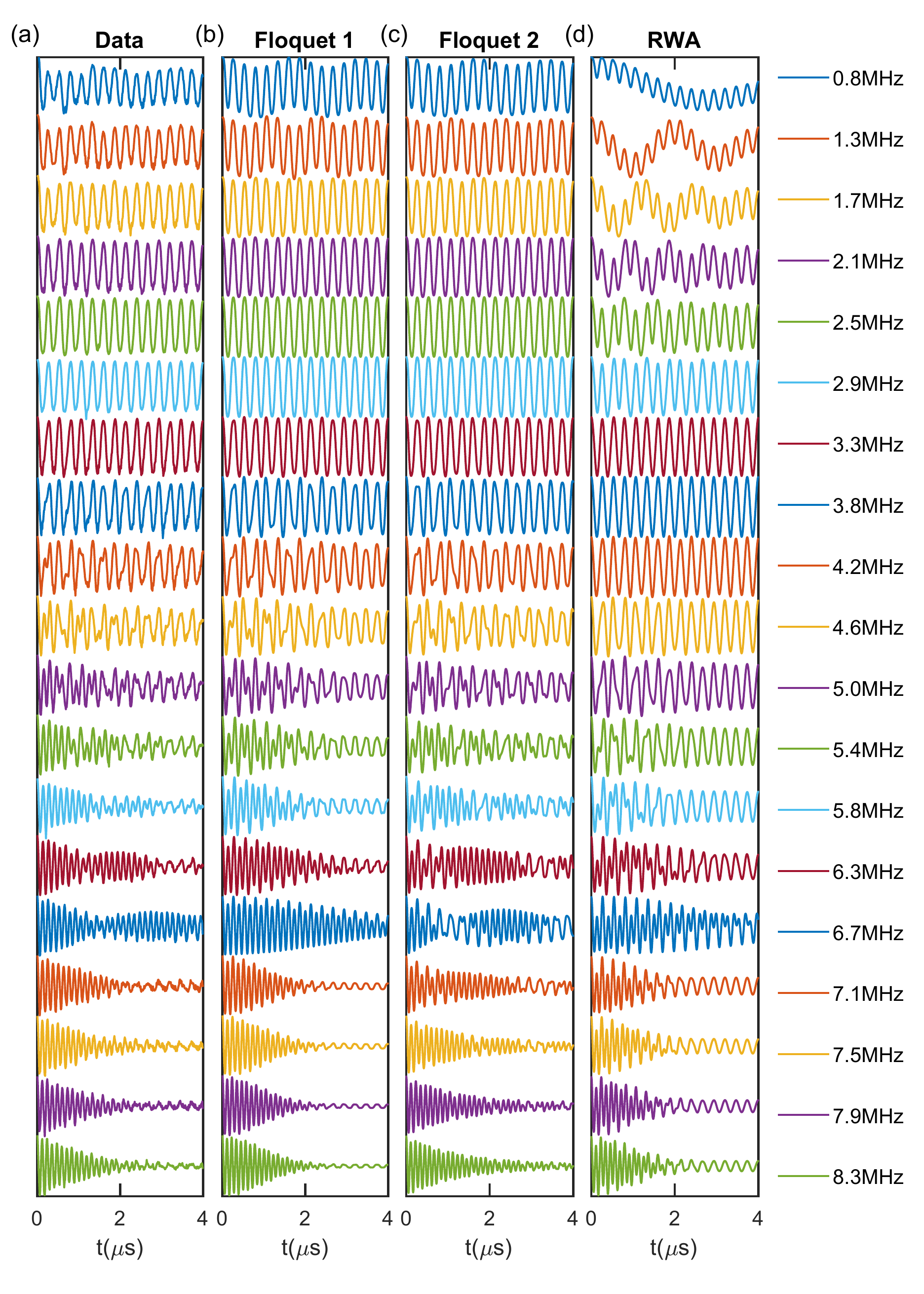}
\caption{\label{RabiCompare_PhiPio2_Decay} Raw data and simulations of Rabi oscillations in the $\Omega$ dependence experiments in main text with $\phi=\pi/2$. The legend is the value of $\Omega/(2\pi)$. (a) Rabi oscillation data. (b) Floquet simulation with decay. Decay factors $\exp(-t/\tau_{i,n})$ are multiplied to each frequency component with $\tau_{i,n}$ fitted from the experiment data. (c) Floquet simulation with decay and avoided crossing by adding an additional term $\epsilon\cos(2\omega_m t)\sigma_y$ to the Hamiltonian in Eq. (1) of the main text and implementing the Floquet simulation. (d) RWA simulation. }
\end{figure*}

\section{Concatenated continuous driving}
Concatenated continuous driving has been explored in several works \cite{caiRobustDynamicalDecoupling2012,khanejaUltraBroadbandNMR2016,saikoSuppressionElectronSpin2018,cohenContinuousDynamicalDecoupling2017,farfurnikExperimentalRealizationTimedependent2017,rohrSynchronizingDynamicsSingle2014,laytonRabiResonanceSpin2014,saikoMultiphotonTransitionsRabi2015,teissierHybridContinuousDynamical2017,bertainaExperimentalProtectionQubit2020,caoProtectingQuantumSpin2020}, typically in the context of protecting a qubit against decoherence. Here we focus on the simplest scheme, where only a second modulation is applied to the first driving field. There are two types of modulation that have been developed:  amplitude modulation or phase modulation. 
By applying an amplitude-modulated microwave along the x axis $\Omega\cos(\omega t)-2\epsilon_m\sin(\omega t)\cos(\omega_m t+\phi)$, the Hamiltonian can be written as
\begin{equation}
    H=\frac{\omega_0}{2}\sigma_z+\left(\Omega\cos(\omega t)-2\epsilon_m\sin(\omega t)\cos(\omega_m t+\phi)\right)\sigma_x
    \label{Hamiltonian}
\end{equation}
where $\omega_0$ is the level splitting of the two-level system, $\Omega,\epsilon_m$ are the driving strengths of the main driving and modulation terms, respectively. 
We will assume $\Omega,\epsilon_m\ll\omega_0$ and $\delta=\omega-\omega_0\ll\omega_0$ throughout this work. In the first rotating frame defined by transformation $U_1=\exp(-i\frac{\omega t}{2}\sigma_z)$, and applying the RWA, the Hamiltonian becomes
\begin{equation}
    H_I=-\frac{\delta}{2}\sigma_z+\frac{\Omega}{2}\sigma_x+\epsilon_m\cos(\omega_mt+\phi)\sigma_y
    \label{HI}
\end{equation} 
In the phase modulation method, the driving waveform has a time-dependent phase, yielding  the Hamiltonian  
\begin{equation}
    H=\frac{\omega_0}{2}\sigma_z+\Omega\cos\left(\omega t+\frac{2\epsilon_m}{\Omega}\cos(\omega_m t+\phi)\right)\sigma_x
    \label{Hamiltonian_PhaseMod}
\end{equation}
We can apply a  rotating frame  transformation $U_1=\exp(-i\int_0^t H_0(t^{'})dt^{'})$ with $H_0(t)=\frac{\omega}{2}\sigma_z- \frac{\epsilon_m\omega_m}{\Omega}\sin(\omega_m t+\phi)\sigma_z$. The transformation is thus given by
$U=\exp\left[-i\left(\frac{\omega t}{2}\sigma_z+\epsilon_m \frac{\cos(\omega_m t+\phi)}{\Omega}\sigma_z \right)\right]$ and within the RWA,  the Hamiltonian in the interaction picture  is
\begin{equation}
        H_I=-\frac{\delta}{2}\sigma_z+\frac{\Omega}{2}\sigma_x+\epsilon_m\frac{\omega_m}{\Omega}\sin(\omega_mt+\phi)\sigma_z
    \label{HI_PhaseMod}
\end{equation}
Note that this Hamiltonian has a form similar to  the Hamiltonian in Eq.~\eqref{HI} obtained with amplitude-modulated driving. However, since the modulation is applied by  varying  the first driving phase, there is no noise associated with $\epsilon_m\frac{\omega_m}{\Omega}$ due to field inhomogeneities or fluctuations (although noise associated with imperfect resolution and faulty electronic elements are still possible.) The phase modulation method  usually has better performance such as longer coherence time \cite{farfurnikExperimentalRealizationTimedependent2017,cohenContinuousDynamicalDecoupling2017}, and less power limitations which enable larger $\epsilon_m$. 
%
\section{Mollow triplet within the rotating wave approximation}
\label{RWA}
The linear oscillating term in Eq.~\eqref{HI} can be decomposed into a co-rotating term $\frac{\epsilon_m}{2}(\cos(\omega_mt+\phi)\sigma_y+\sin(\omega_mt+\phi)\sigma_{z^{'}})$ and a counter-rotating term $\frac{\epsilon_m}{2}(\cos(\omega_mt+\phi)\sigma_y-\sin(\omega_mt+\phi)\sigma_{z^{'}})$, where $\sigma_{z^{'}} = \cos\beta \sigma_z +\sin\beta \sigma_x$, with $\sin\beta = \frac{\delta}{\sqrt{\Omega^2+\delta^2}}$. By dropping the counter-rotating terms, in the second rotating frame defined by $-\frac{\omega_m}{2}\sigma_{x^{'}}$ where $\sigma_{x^{'}} = \cos\beta \sigma_x -\sin\beta \sigma_z$, Hamiltonian becomes $H_I^{(2)}=\frac{1}{2}(\sqrt{\delta^2+\Omega^2}-\omega_m)\sigma_{x^{'}}+\frac{\epsilon_m}{2}(\cos\phi\, \sigma_y +\sin \phi\,\sigma_{z^{'}})$. The spin evolution in the second rotating frame is simply $|\psi(t)\rangle_I^{(2)} =e^{-i H_I^{(2)}t} |\psi(0)\rangle$. In the first rotating frame, the spin state is $|\psi(t)\rangle_I^{(1)}=e^{-i \frac{\omega_m t}{2}(-\sin \beta\,\sigma_z+\cos \beta \, \sigma_x)}|\psi(t)\rangle_I^{(2)}$. Going back to the lab frame adds an additional relative phase between $|0\rangle$ and $|1\rangle$ but keeps the population unchanged. Such procedure predicts that the population in $|0\rangle$ is a sum over the three frequency components of the Mollow triplet, $\omega_m,\omega_m+\sqrt{\epsilon_m^2+(\omega_m-\Omega_R)^2}, \omega_m-\sqrt{\epsilon_m^2+(\omega_m-\Omega_R)^2}$ where $\Omega_R=\sqrt{\Omega^2+\delta^2}$ is the effective Rabi frequency. Compared with the case of normal Rabi oscillation, the population measurement in the CCD scheme does not commute with the second rotating frame, making it possible to observe the Mollow triplet \cite{rohrSynchronizingDynamicsSingle2014,teissierHybridContinuousDynamical2017,pigeauObservationPhononicMollow2015}.

\section{Avoided Crossing and power saturation}

We compare amplitude-modulated (this supplement) and phase-modulated (main text) CCD scheme in the $\epsilon_m$ dependence experiments. The amplitude-modulated experiments show additional avoided crossing features and non-vanishing components. In Figs.~\ref{emsweep_dataSim}(a) and \ref{emsweep_dataSim}(f), the FFT of amplitude modulation has avoided crossing features when the frequency components $\lambda^+-\lambda^-, 2\omega_m-(\lambda^+-\lambda^-)$ cross, and this is caused by the power saturation. By adding an additional $\epsilon\cos(2\omega_m t)$ to the $H_I$ in Eq. (1) in the main text and performing the Floquet calculation, we can reproduce such avoided crossing and the non-vanishing $\lambda^+-\lambda-$ components in Figs.~\ref{emsweep_dataSim}(d) and \ref{emsweep_dataSim}(i). As a reference, Figs.~\ref{emsweep_dataSim}(e) and \ref{emsweep_dataSim}(j) are simulations with no such additional terms. In addition, to further explore these features, we perform similar experiments but scale down the driving strength $\Omega,\omega_m,\epsilon_m$ by a half in Figs.~\ref{emsweep_dataSim}(b) and \ref{emsweep_dataSim}(g) and a quarter in Figs.~\ref{emsweep_dataSim}(c) and \ref{emsweep_dataSim}(h). In comparison, the relative splitting of energy levels and the intensity in the region of the avoided crossing around $\epsilon_m\approx\Omega$ are smaller when driving strength is weaker. With smaller power, we are able to measure to higher $\epsilon_m/\Omega$. In Figs.~\ref{emsweep_dataSim}(b) and \ref{emsweep_dataSim}(g), there is another avoided crossing measured at $\epsilon_m\sim (2\pi)3.5\text{MHz}$ which is caused by the mixing of the same frequency components as the crossing at $\epsilon_m\sim \Omega$, and this avoided crossing is more prominent than that seen at $\epsilon_m\sim \Omega$, which is another piece of evidence that the avoided crossing is caused by the power saturation.

Figure~\ref{RabiFreqAmp_Saturation}(a) shows that the power saturates when the Rabi frequency approaches $(2\pi)7\sim (2\pi)9$MHz. And in Fig.~\ref{RabiFreqAmp_Saturation}(c) we plot how the peak-to-peak amplitude (blue circles) and the average amplitude (blue triangles) of the waveform depend on the setting value $\Omega$. When $\epsilon_m$ approaches $(2\pi)3\sim (2\pi)4$MHz, the maximum amplitude of the amplitude-modulated driving waveform exceeds the saturation level although the average amplitude is still not saturated. In Fig.~\ref{RabiSaturationFFT}(b) we simulate the amplitude-modulated waveform with a saturation level of the output voltage setting to $8$MHz. In the FFT analysis of the simulated waveform, higher order frequency components such as $2\omega_m,3\omega_m$ start to appear in Fig.~\ref{RabiSaturationFFT}(f) for the amplitude-modulated case. 

Returning to the comparison between amplitude modulation and phase modulation, we show above that the amplitude modulation is limited by the power saturation of the microwave delivery. As for the phase modulation, frequency expansion of the phase-modulated waveform $\Omega\cos(\omega_0+\frac{2\epsilon_m}{\Omega}\cos(\omega_m t+\phi))$ includes an infinite series of frequency components $\omega_0,\omega_0\pm\omega_m,\omega_0\pm 2\omega_m,\cdots$. Larger $\frac{2\epsilon_m}{\Omega}$ results in more prominent side bands. We compare a strong modulation in Figs.~\ref{RabiSaturationFFT}(c) and \ref{RabiSaturationFFT}(g) and weak modulation in Figs.~\ref{RabiSaturationFFT}(d) and \ref{RabiSaturationFFT}(h). The FFT spectra of strong modulation shows more prominent side bands. The region between the dashed lines is the working range of our electronics elements. As a result, the phase modulation is limited by the range of electronic elements. 

\section{Raw data for $\Omega$ dependence experiments}

The following figures are raw data and simulations of the experiments in the main text to show the resonance shift by sweeping $\Omega$. Figure~\ref{RabiCompare_Phi0} and Fig.~\ref{RabiCompare_PhiPio2} are the comparisons between data (a) and simulation (b,c,d) of Rabi oscillations under different $\Omega$ when $\phi=0$. Simulations in (b) are calculated by directly evolving the Hamiltonian. (c) and (d) are calculated by summing over the frequency components in the Floquet calculation without and with the RWA correspondingly. The consistency of (b) and (c) serves as a verification that the Floquet approach is a precise way to describe the dynamics of the system. The data shows large differences between (c) and (d) when $\Omega$ is small, which indicates that when the ratio of $\epsilon_m$ to $\Omega$ becomes large, the counter-rotating effects start to appear. To further compare the data with simulation, we add decay factors $\exp(-t/\tau_{i,n})$ with $\tau_{i,n}$ fitted from the Rabi oscillations measured in (a) to the simulations in (c) and (d) and plot the comparisons in Fig.~\ref{RabiCompare_Phi0_Decay} and Fig.~\ref{RabiCompare_PhiPio2_Decay}. The data shows good consistency with the Floquet simulation in (c) where we add the additional modulation term $\epsilon\cos(2\omega_m t)\sigma_y$ to the Hamiltonian in Eq. (1) of the main text. As a comparison, the simulation in (b) is without such additional term. Thus the beating in the oscillations when $\Omega/(2\pi)=6\sim7\text{MHz}$ is the result of avoided crossing caused by the power saturation in the amplitude modulated driving waveform which generates the mixing between $\omega_m+\lambda^+-\lambda^-, 3\omega_m-(\lambda^+-\lambda^-)$. Since the power saturation is minuscule in this experiment as shown in Fig.~\ref{RabiFreqAmp_Saturation}(c) in red points, the avoided crossing is almost not visible in FFT spectra in the main text and can be clearly seen as the gradual switch of two frequency components $\omega_m+\lambda^+-\lambda^-$ and $3\omega_m-(\lambda^+-\lambda^-)$ as discussed in the main text and Appendix. 

\end{document}